\documentclass{aa}  
\usepackage{xcolor}
\usepackage{graphicx}
\usepackage{longtable}
\usepackage{amsmath}
\usepackage{mwe}
\usepackage{fontawesome}

\definecolor{dgreen}{rgb}{0.0, 0.5, 0.0}
\usepackage{txfonts}
\usepackage{hyperref}
\hypersetup{
    colorlinks=true,
    linkcolor=blue,
    filecolor=blue,
    citecolor=blue,
    urlcolor=blue,
}

\begin{document}

   \title{Water ice: temperature-dependent refractive indexes and their astrophysical implications}

   \author{W. R. M. Rocha\inst{1,2},
          M. G. Rachid\inst{1},
          M. K. McClure\inst{2},
          J. He\inst{3},
          \and
          H. Linnartz\inst{1}
          }

   \institute{Laboratory for Astrophysics, Leiden Observatory, Leiden University, P.O. Box 9513, NL 2300 RA Leiden, The Netherlands.\\
    \email{rocha@strw.leidenuniv.nl}
          \and
             Leiden Observatory, Leiden University, PO Box 9513, NL 2300 RA Leiden, The Netherlands
         \and
             Max Planck Institute for Astronomy, K{\"o}nigstuhl 17, D-69117 Heidelberg, Germany
             }

   \date{Received xxxx; accepted yyyy}

 
  \abstract
   {Inter- and circumstellar ices are largely composed of frozen water. Therefore, it is important to derive fundamental parameters for H$_2$O ice such as absorption and scattering opacities for which accurate complex refractive indexes are needed.}
   {The primary goal of the work presented here is to derive ice-grain opacities based on accurate H$_2$O ice complex refractive indexes at low temperatures and to assess the impact this has on the derivation of ice column densities and porosity in space.}
   {We use the \texttt{optool} code to derive ice-grain scattering and absorption opacity values based on new and previously reported mid-IR complex refractive index measurements of H$_2$O ice, primarily in its amorphous form, but not exclusively. Next, we use those opacities in the \texttt{RADMC-3D} code to run a radiative transfer simulation of a protostellar envelope containing H$_2$O ice, which is then used to calculate water ice column densities.}
   {We find that the real refractive index in the mid-IR of H$_2$O ice at 30~K is $\sim$14\% lower than previously reported in the literature. This has a direct impact on the ice column densities derived from the simulations of embedded protostars. Additionally, we find that ice porosity plays a significant role in the opacity of icy grains and that the H$_2$O libration mode can be used as a diagnostic tool to constrain the porosity level. Finally, the refractive indexes presented here allow us to estimate a grain size detection limit of 18~$\mu$m based on the 3~$\mu$m band whereas the 6~$\mu$m band allows tracing grain sizes larger than 20~$\mu$m.}
   {Based on radiative transfer simulations using new mid-IR refractive indexes, we conclude that H$_2$O ice leads to more absorption of infrared light than previously estimated. This implies that the 3 and 6~$\mu$m bands remain detectable in icy grains with sizes larger than 10~$\mu$m. Finally, we propose that also the H$_2$O ice libration band can be used as a diagnostic tool to constrain the porosity level of the interstellar ice, in addition to the OH dangling bond, which is now routinely used for this purpose.}

\authorrunning{Rocha et al.}
\titlerunning{water ice refractive index}
\maketitle
%

\section{Introduction}
Water is ubiquitously found on Earth in its different physical states, i.e., gas, liquid and solid. Frozen water has also been found on different moons in the solar system \citep[e.g.,][]{Encrenaz2008, Bell2010, Snodgrass2017}. Outside our solar system, in star-forming regions, water has been identified in the gas phase \citep[e.g.,][]{van_Dishoeck2021} and in the solid phase. The first detection of water ice in the interstellar medium (ISM) was made by \citet{Gillett1973}. Different space telescope missions corroborated this detection, such as the {\it Infrared Space Observatory} \citep[ISO, e.g.,][]{Gibb2000}, {\it Spitzer} Space Telescope \citep[e.g.,][]{Pontoppidan2005}, and more recently, with the {\it James Webb} Space Telescope \citep[JWST, e.g.,][]{Yang2022, McClure2023, Beuther2023}. H$_2$O ice has also been observed by ground-based telescopes located at high altitudes, e.g., the {\it Very Large Telescope} \citep[VLT; e.g.,][]{Pontoppidan2004}, the Keck observatory \citep[e.g.,][]{Boogert2008}, and the Infrared Telescope Facility \citep[IRTF, e.g.,][]{Chu2020}. 

Water is the major component of ice mantles on dust grains, and traces dense regions ($n > 10^3$~cm$^{-3}$) with an estimated visual extinction threshold ($A_{\rm{V}}$) of 1.6~mag \citep{Whittet2001, Boogert2015}. As the dominant component in inter- and circumstellar ices, water ice also significantly regulates the absorption and scattering of irradiation in specific wavelengths. Because of the latter, it is crucial to quantify accurate physicochemical parameters of H$_2$O ice, specifically, the temperature and wavelength-dependent complex refractive index (CRI). These values are the starting point for the work presented here.

Experimentally, the derivation of such CRI H$_2$O parameters is challenging, because of the generally limited spectral coverage of the used techniques and the dependence of ice structure (i.e., level of porosity) on temperature. Additionally, deriving refractive indexes in the IR requires knowledge of the refractive index in the visible range, typically around $\sim$0.7~$\mu$m. Until recently, these values were not available from experiments targeting cold interstellar ice conditions. For example, \citet{Irvine1968} compiled and discussed measurements of the water ice refractive index back in 1936. In most cases, the experiments were performed at temperatures above 80~K, which is substantially warmer than the typical temperatures inside molecular clouds. More refractive index values were proposed by \citet{Hale1973}, who performed an experiment at $\sim$300~K. In the values compiled by \citet{Irvine1968} and measured by \citet{Hale1973}, the refractive index of H$_2$O crystalline ice was found to be around 1.32. This number was used by \citet{Hudgins1993} to calculate the CRI values for H$_2$O in the range from 2.5 to 200~$\mu$m and for temperatures in the range from 10 to 140~K. New measurements for water ice at 22~K were performed by \citet{Dohnalek2003}, who obtained a refractive index of the fully compact amorphous H$_2$O ice equal to 1.29 at 0.63~$\mu$m. Using this new refractive index, \citet{Mastrapa2009} calculated the H$_2$O ice CRI values starting at 1.4~$\mu$m and for low temperatures. However, in recent studies focusing on the UV-visible range, \citet{Kofman2019} and \citet{He2022} showed that the UV-vis refractive index of amorphous H$_2$O ice around 30~K is lower than proposed by \citet{Dohnalek2003}, or those used by \citet{Hudgins1993}. \citet{Stubbing2020} also derived experimental refractive indexes of H$_2$O ice, using a different conceptual approach, that is in good agreement with the amorphous H$_2$O ice values derived by \citet{Kofman2019}. This approach measures the ice refractive index after the ice deposition, instead of during the ice condensation, as is more common. This allows for probing variations in the ice structure due to thermal or energetic processing. The approach also allows to provide imaginary refractive index of H$_2$O ice, which is not the case in other methods. Motivated by these recent refractive indexes derived in the UV-vis range, in particular, the latest values presented in \citet{He2022}, new mid-IR CRI values were calculated by \citet{Rocha2022}, which are publicly available in the Leiden Ice Database for Astrochemistry \footnote{\url{https://icedb.strw.leidenuniv.nl/}} (LIDA).

CRI values are crucial information to perform radiative transfer simulations in dusty media, and therefore, providing accurate values is of utmost importance for interpreting astronomical data. Moreover, with the advent of the JWST era, these data are urgently needed to support the analysis of ices in star-forming regions. For this reason, we exploit here the new CRI values from \citet{Rocha2022} as input for ice-grain opacity values and assess their impact on astrophysical observables, such as the ice column density. Extending on this, we also discuss how the most recent water ice refractive indexes at low temperatures contribute to elucidating the size limit of large grains where water ice is no longer detectable.

This paper is organized as follows. In Section 2 we provide an overview of the CRI values available in LIDA. Section~3 shows the methodology adopted to derive ice-grain opacity values and to perform the radiative transfer simulations. Section 4 presents the results of this work, and in Section~5 we discuss the astrophysical implications of the newly derived CRI values. Finally, the conclusions are presented in Section 6.


    






\section{Overview of the refractive indexes in LIDA}
\label{fromlida}
 In this section, we provide an overview of the experimental method used to calculate the temperature-dependent water ice refractive indices shown in \citet{Rocha2022} for a spectral range between 0.3 and 20~$\mu$m and from 30 to 135~K, and how these compare with previous values in the literature. Additionally, we calculate the CRI values of crystalline H$_2$O ice at 160~K.

\subsection{Range between 0.3 and 0.7~$\mu$m}
\label{range1}
The water ice refractive index between 0.3 and 0.7~$\mu$m (300$-$700~nm) is measured using the Optical Absorption Setup for Ice Spectroscopy (OASIS) in the Laboratory for Astrophysics at Leiden Observatory. The experimental method used is detailed in \citet{He2022} and only briefly commented on here.

A high-vacuum chamber with a base pressure of 10$^{-8}$~mbar is used to perform the experiments. Pure water ice grows {\it in-situ} via background deposition onto a cryogenically cooled  UV-enhanced aluminium mirror (Thorlabs Inc.). The deposition rate is manually controlled and the chamber pressure during this procedure is around 5$\times$10$^{-5}$~mbar. The water vapour is deposited at different temperatures, i.e., from 30 to 150~K. During the deposition, two light sources illuminate the same ice surface simultaneously. One source is a frequency-stabilized HeNe laser at a wavelength of 632.8~nm, and the second source is a broadband Xe-arc lamp, covering 0.3 to 0.7~$\mu$m. When one of these beams reaches the growing ice, part of the light is diffracted into the ice film and is reflected by the aluminium mirror. The HeNe laser beam is reflected towards a photodiode detector and the Xe-arc lamp is reflected towards a commercial UV-vis spectrometer, where the light is spectrally dispersed and the signals are recorded using a CCD camera. Another part of the HeNe or Xe-arc beam is reflected already by the ice surface and detected in the same way. The optical path difference between these two beams creates an interference pattern, and for the broadband source, this is observed for all wavelengths. The fitting of the interference pattern allows to calculate the period of oscillation; this information is then used to derive the ice refractive index using Equation~\ref{n_uvvis}:
\begin{equation}
    n_{uv-vis} = \sqrt{\frac{\sin^2 \alpha-(P_{\beta}/P_{\alpha})^2 \sin^2 \beta}{1 - (P_{\beta}/P_{\alpha})^2}},
\label{n_uvvis}
\end{equation}
where the periods of the interference patterns generated by the two light beams striking the ice are $P_{\rm{\alpha}}$ and $P_{\rm{\beta}}$, and $\alpha$ and $\beta$ are the corresponding angles between the normal plane and the light beam. In our experiment, these are 45$^{\circ}\pm$4$^{\circ}$ for the Xe-arc lamp light and 4.4$^{\circ}$ for the laser, respectively.

\begin{figure}
   \centering
   \includegraphics[width=\hsize]{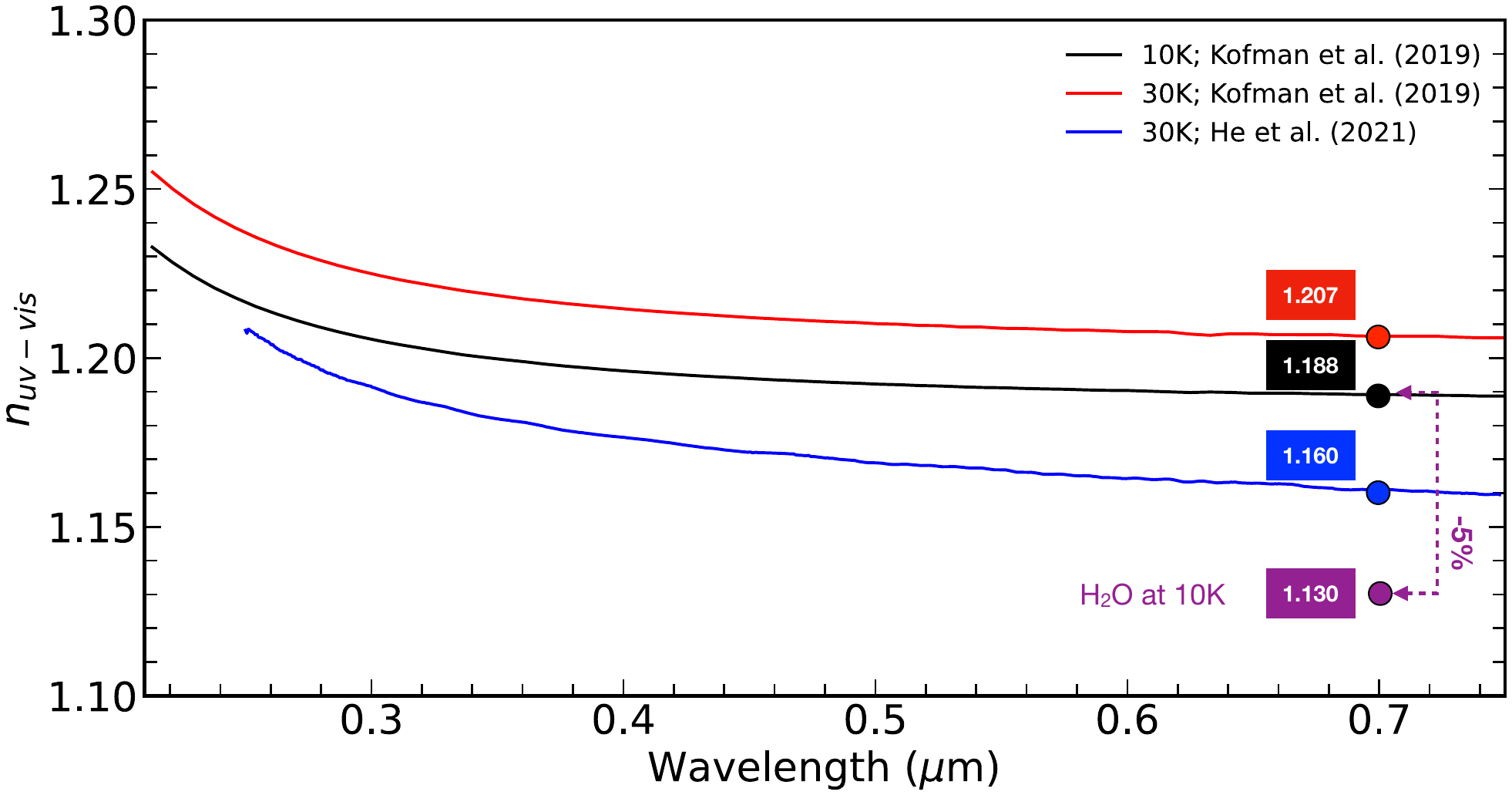}
      \caption{Comparison between $n_{uv-vis}$ values derived by \citet{Kofman2019} and \citet{He2022}. The refractive index values at 0.7~$\mu$m are indicated in the figure. The purple circle and box show the $n_{uv-vis}$ from \citet{Kofman2019} at 10~K reduced by 5\%.}
         \label{n_vals}
   \end{figure}

Figure~\ref{n_vals} compares the wavelength-dependent water ice refractive index measured by \citet{Kofman2019} and \citet{He2022}. This graph shows that the values derived by \citet{He2022} at 30~K are lower than those obtained from \citet{Kofman2019}. This difference is not due to an experimental artefact, but the result of the method used by \citet{Kofman2019} that requires an estimated H$_2$O ice density, similar to the technique employed by \citet{Dohnalek2003}. Instead, the method used by \citet{He2022} is intrinsically independent of the ice density, which allows us to derive more precise values. At 0.7~$\mu$m and 30~K, the difference between the values from \citet{Kofman2019} and \citet{He2022} is 5\%. The figure also shows the refractive index at 10~K from \citet{Kofman2019}; the $n_{uv-vis}$ values are again lower than refractive index values from \citet{Dohnalek2003}. Moreover, by comparing \citet{Kofman2019} and \citet{He2022} values at 30~K, it is possible to see a different decreasing pattern from short to long wavelengths. As for 10~K, no corresponding values exist in \citet{He2022}, and another direct comparison is not possible. However, assuming that the difference of 5\% found for $n_{uv-vis}$ between \citet{Kofman2019} and \citet{He2022} at 30~K and at 0.7~$\mu$m is also valid for 10~K, we estimate the $n_{uv-vis}$ value for H$_2$O ice to be 1.13, instead of 1.19 as reported in \citet{Kofman2019}.



\subsection{Range between 2 and 20~$\mu$m}
\label{range2}
The water ice refractive index between 2 and 20~$\mu$m is derived using the online refractive index calculator\footnote{\url{https://icedb.strw.leidenuniv.nl/Kramers_Kronig}} available through LIDA. This tool calculates the real ($n$) and imaginary ($k$) parts of the complex refractive index ($\Tilde{m} = n + ik$) from the absorbance H$_2$O ice spectrum. More details about the formalism involved in the calculations can be found in \citet{Hudgins1993, Rocha2014, Gerakines2020} and \citet{Mvondo2022}.

We use the pure H$_2$O absorbance spectrum from LIDA\footnote{\url{https://icedb.strw.leidenuniv.nl/data/13}}. The used spectra of amorphous H$_2$O ice ($30~\mathrm{K} \leq T \leq 135~\mathrm{K}$) were measured by \citet{Oberg2007}. In the experiments, water vapour was deposited onto a CsI substrate at specific temperatures. Next, transmission IR spectroscopy was used to record the spectrum between 2 and 20~$\mu$m and at a resolution of 2 cm$^{-1}$.

The methodology consists of deriving the $k$ values first, which is given by:
\begin{equation}
    k = \frac{1}{4\pi \nu d} \cdot \left( 2.3 \times Abs_{\nu} + \mathrm{ln} \left| \frac{\tilde{t}_{01}\tilde{t}_{02}}{1 + \tilde{r}_{01} \tilde{r}_{12} e^{2i\tilde{x}}} \right|^2 \right)
    \label{k-value}
\end{equation}
where $Abs_{\nu}$ is the absorbance spectrum, and $\nu$ is the wavenumber corresponding to the peak position of the band, $d$ is the thickness of the ice, and $\tilde{t}_{01}$, $\tilde{t}_{02}$, $\tilde{r}_{01}$, $\tilde{r}_{12}$ are the Fresnel coefficients, which account for the amount of light crossing the ice and reaching the substrate, as well as reflected by the substrate or the ice surface. The numbers 0, 1 and 2 refer to regions of vacuum, ice sample, and substrate, respectively. The refractive index of the substrate is implicit in the terms $\tilde{t}_{02}$ and $\tilde{r}_{12}$. Finally, the term $\tilde{x}$ is given by $\tilde{x} = 2\pi \nu d \tilde{m}$. When the $k$ values are known for each frequency $\nu$, the $n$ values are calculated using the Kramers-Kronig relation shown below:
\begin{equation}
    n(\nu) = n_{700nm} + \frac{2}{\pi} \mathcal{P} \int_{\nu_1}^{\nu_2} \frac{\nu' k(\nu')}{\nu'^2 - \nu^2}d\nu'
    \label{n-value}
\end{equation}
where $n_{\mathrm{700nm}}$ is the refractive index of the sample at 700~nm, i.e., where the UV-vis spectral coverage ends, $\nu'$ is the wavenumber around the $\nu$ value. The Cauchy principal value $\mathcal{P}$ is used to overcome the singularity when $\nu = \nu'$.

The online tool in LIDA solves Equations~\ref{k-value} and \ref{n-value} iteratively, i.e., the $n$ value at each iteration allows us to refine the calculation of the Fresnel coefficients. As a result, the improved $k$ values result in an improved $n$ value. The procedure continues until the variation in subsequent $n$ and $k$ values is less than 0.1\%. Before running this online tool, a few input parameters are requested, which are the ice thickness, the refractive index at 700~nm, the refractive index of the substrate and the mean average percentage error, MAPE (Mean Average Percentage Error). Table~\ref{nkabs_par} lists the input data used to calculate the water ice refractive index between 2 and 20~$\mu$m, the number of iterations used by the tool and the final MAPE.

\begin{table}
\caption{Input parameters used to calculate the mid-IR water ice refractive index (2$-$20~$\mu$m).}
\label{nkabs_par}      
\centering
\setlength{\tabcolsep}{3pt} 
\renewcommand{\arraystretch}{1.5} 
\begin{tabular}{l c c}        
\hline\hline                 
Parameters & Values & Reference\\
\hline
\multicolumn{3}{c}{\it{IR spectrum from \citet{Oberg2007}; 15$-$135~K}}\\
\hline
Thickness (cm) & 1 $\times$ 10$^{-4}$ & \citet{Oberg2007}\\
$n_{700nm}^{30K}$ & 1.16 & \citet{He2022}\\
$n_{700nm}^{75K}$ & 1.21 & \citet{He2022}\\
$n_{700nm}^{105K}$ & 1.23 & \citet{He2022}\\
$n_{700nm}^{135K}$ & 1.25 & \citet{He2022}\\
\hline
\multicolumn{3}{c}{\it{IR spectrum from \citet{Gerakines1996}; 160~K}}\\
\hline
Thickness (cm) & 0.3 $\times$ 10$^{-4}$ & \citet{Gerakines1996}\\
$n_{700nm}^{160K}$ & 1.31 & \citet{He2022}\\
\hline
\multicolumn{3}{c}{\it{CsI substrate}}\\
\hline
$n_{\rm{substrate}}$ & 1.73 & \citet{Querry1987}\\
\hline
\multicolumn{3}{c}{\it{Calculation}}\\
\hline
Iterations & 5 & ...\\
MAPE & $\leq$ 0.1\% & ...\\
\hline
\end{tabular}
\end{table}

\subsection{Joining UV-vis and mid-IR data}
\label{joining}
The data described in Sections~\ref{range1} and \ref{range2} covers two spectral ranges, i.e., 0.3 to 0.7~$\mu$m and 2 to 20~$\mu$m. Clearly, for the interval between 0.7 and 2~$\mu$m, experimental spectra are not available. To cover this range the two data sets need to be linked, and we use different approaches for $n$ and $k$. For $k$ values we extrapolate the imaginary refractive index at 2~$\mu$m (10$^{-4}$) until 0.3~$\mu$m. In the case of $n$, we use a low order polynomial to link the water ice $n$ values from \citet{He2022} to the data starting at 2~$\mu$m. The caveat in this approach is that we have to neglect the water ice absorption bands in the interval between 0.7 and 2~$\mu$m. In this range, water ice has overtone transitions between 1.4 and 1.8~$\mu$m, but they are weak, both for amorphous and crystalline ice \citep{Mastrapa2008}. For example, the $k$ values calculated by \citet{Mastrapa2008} range from 10$^{-5}$ to 10$^{-3}$ which covers the value used in our extrapolation (10$^{-4}$). Similarly, the 
variation in $n$ is rather small.

\subsection{CRI values at 160~K}
In addition to the CRI values of amorphous water ice taken from LIDA, and shown in Section~\ref{range2}, we add the CRI values for crystalline ice following the steps described in Sections~\ref{range1}-\ref{joining}. For this, we used the infrared spectrum of crystalline H$_2$O ice ($T = 160~\rm{K}$) measured by \citet{Gerakines1996}, who follows the same experimental method described in Section~\ref{fromlida}. Table~\ref{nkabs_par} lists the parameters used in the calculations.

\subsection{UV to mid-IR refractive index: 0.3$-$20~$\mu$m}

\begin{figure}
   \centering
   \includegraphics[width=9cm]{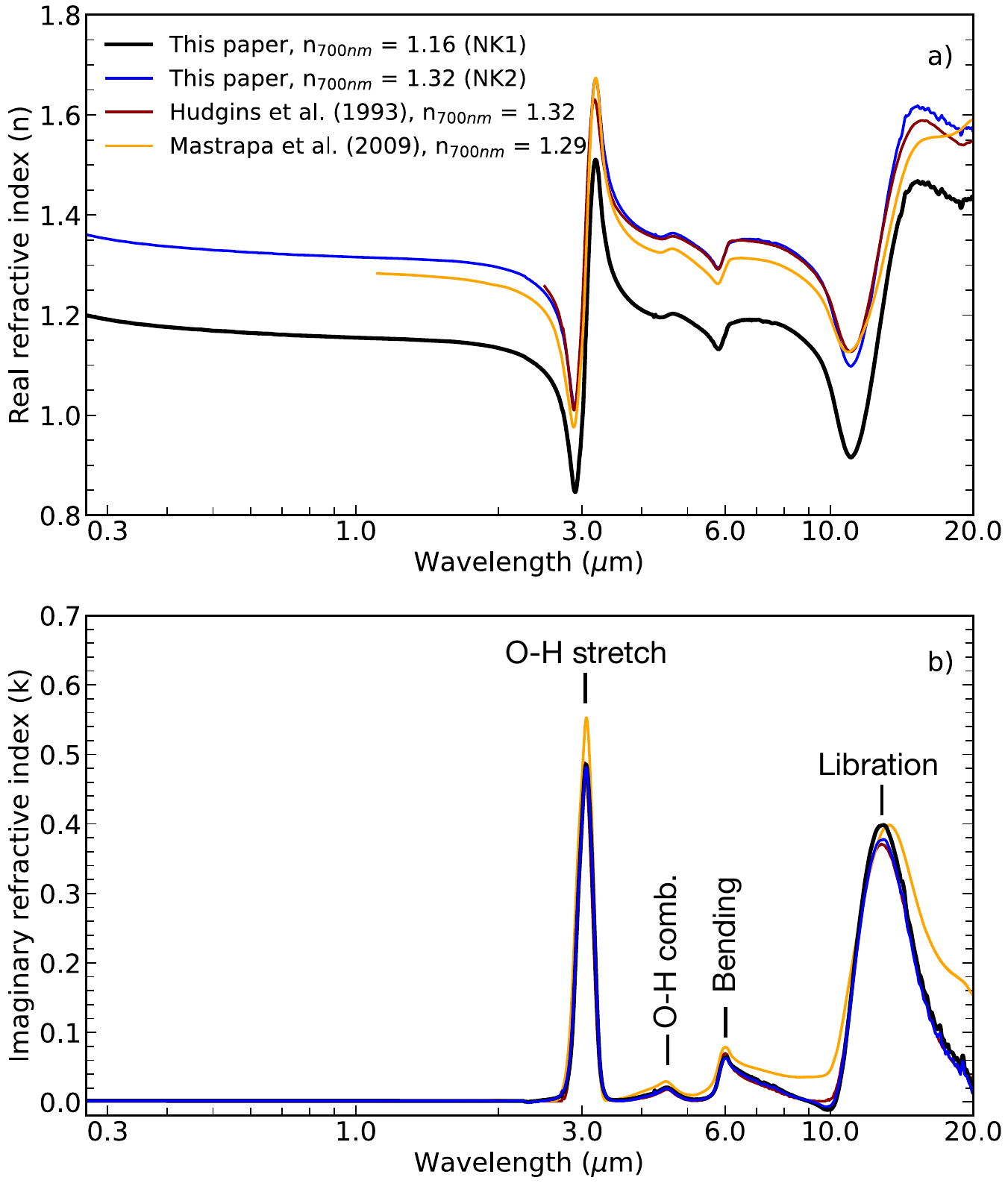}
      \caption{Wavelength-dependent refractive index of H$_2$O ice. Panels {\it a} and {\it b} show the real and imaginary parts of the complex refractive index derived in this paper (H$_2$O at 30~K) adopting $n_{700nm} = 1.16$ (black) and $n_{700nm} = 1.32$ (blue) compared to literature values from \citet{Hudgins1993} given by the purple line (H$_2$O at 40~K), and from \citet{Mastrapa2009}, shown by the orange line (H$_2$O at 25~K). The inset in panel {\it b} displays an illustration of the water vibrational modes and their corresponding motion, except the O$-$H combination mode at 4.5~$\mu$m. Each mode is indicated at the top of the bands. The molecules cartoon was adapted from \url{https://water.lsbu.ac.uk/water/water_vibrational_spectrum.html}.}
         \label{allN}
   \end{figure}

In Figure~\ref{allN}a and \ref{allN}b we show, respectively, the real and imaginary refractive index of pure amorphous water ice ranging from 0.3 to 20~$\mu$m at 30~K. We compare our results with previous data from \citet{Hudgins1993} at 40~K and from \citet{Mastrapa2009} at 25~K. Four water bands are indicated in panel b, which correspond to the symmetric and asymmetric O$-$H stretching mode at 3~$\mu$m ($\nu_1$ and $\nu_3$), the bending mode at 6~$\mu$m ($\nu_2$) and the libration mode, also known as hindered rotation, at 13~$\mu$m ($\nu_L$). The band at 4.5~$\mu$m is due to the combination of the bending and libration modes ($\nu_2$ + $\nu_L$). For the 30~K measurements presented here, a larger spectral coverage is achieved because of the used broadband method. This ensures more accuracy when deriving physical parameters depending on the refractive index.

To derive the CRI values of water ice at 30~K, we use $n_{700nm} = 1.16$ from \citet{He2022}, which is labelled as NK1 in Figure ~\ref{allN}a. For comparison with the literature values, we perform the same calculation, by assuming $n_{700nm} = 1.32$ as adopted by \citet{Hudgins1993}. Next, we shifted up the UV/vis data at 30~K to account for the difference in the $n_{700nm}$ value. The $n$ and $k$ values calculated from this second approach are labelled as NK2. As one can note, the NK2 CRI values are similar to what is derived by \citet{Hudgins1993}, which shows that our numerical approach is consistent with \citet{Hudgins1993}. The difference between NK1 and NK2 values is 13\%. This difference is lower if compared to \citet{Mastrapa2009}, who adopted $n_{700nm} = 1.29$ taken from \citet{Dohnalek2003}, measured at 22~K and assuming compact ice. In Figure~\ref{allN}b, we show the $k$ values. There is a better agreement between the band shapes derived in the present paper and \citet{Hudgins1993}. However, a small difference ($\sim$7\%) is seen in the peak intensity at 13~$\mu$m. The major differences in the peak intensities are observed in the spectrum from \citet{Mastrapa2009}, and the water libration band is also broader. One can also note that the differences in the $n$ values do not necessarily imply larger differences in the $k$ values as also pointed out by \citet{Mastrapa2008}. They show that the $k$ values are much less sensitive to the $n_0$ values, which is also observed here.

\begin{figure*}
   \centering
   \includegraphics[width=\hsize]{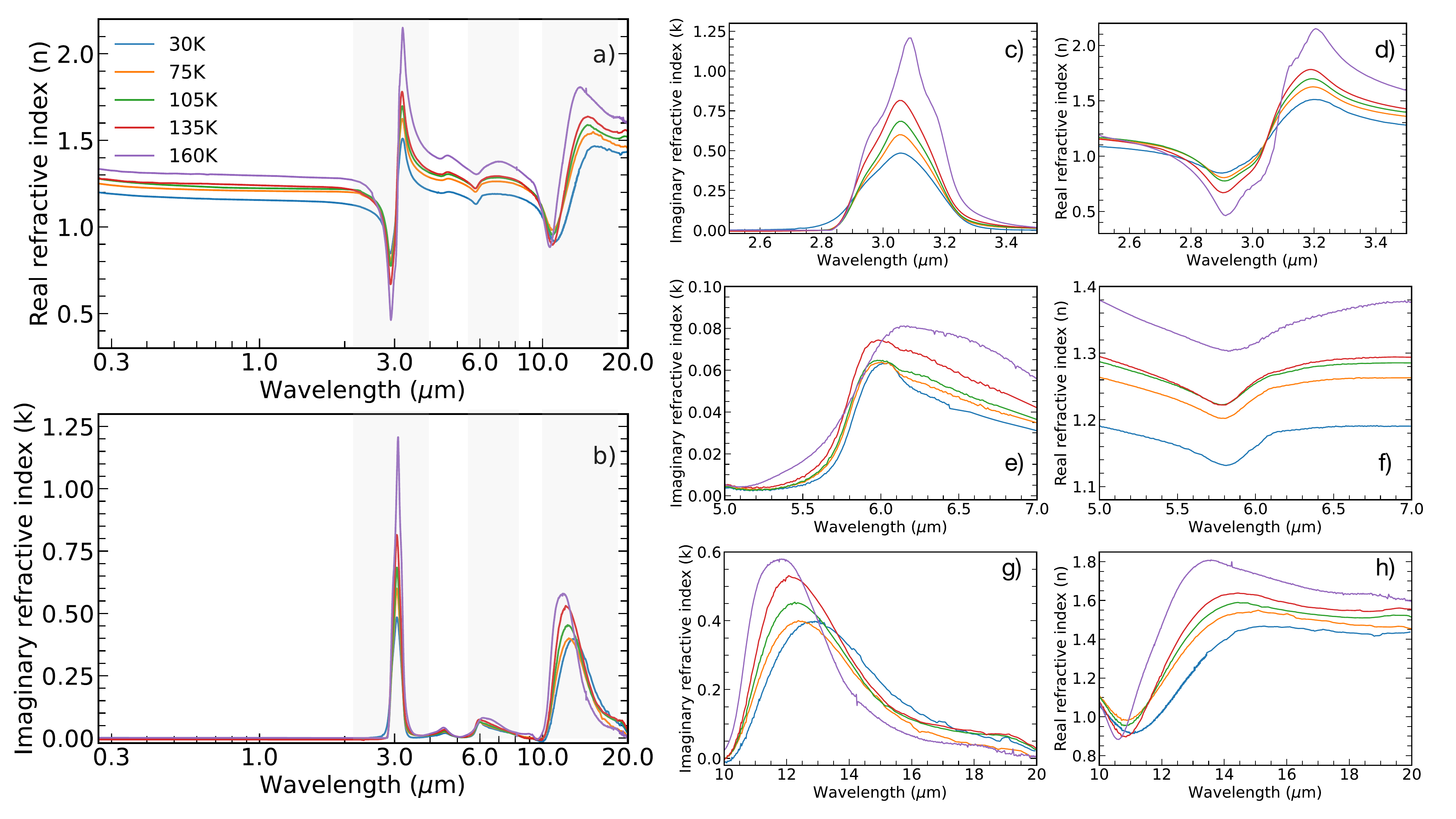}
      \caption{Wavelength-dependent refractive index of H$_2$O ice at different temperatures. Panels {\it a} and {\it b} show the real and imaginary parts of the complex refractive index in a broad-range perspective (0.3$-$20~$\mu$m). Panels {\it c$-$h} show zoom-ins of selected wavelengths in panels {\it a} and {\it b} indicated by the hatched areas.}
         \label{nkallT}
   \end{figure*}

Figure~\ref{nkallT} shows the real and imaginary parts of the H$_2$O ice at temperatures of 30, 75, 105, and 135~K taken from LIDA and at 160~K derived in this work. The vertical offset in the real refractive indexes displayed in Fig. \ref{nkallT}a is due to the different $n_{700nm}$ values used in the calculations (see Table~\ref{nkabs_par}). Furthermore, one can note differences in the band intensities of the imaginary refractive index at 3~$\mu$m and $\sim$12~$\mu$m (Fig. \ref{nkallT}b) and a blue-shift of the bands for increasing temperatures. Figures~\ref{nkallT}c$-$h show details of the $n$ and $k$ profiles in selected spectral ranges. The sharpening of the 3~$\mu$m band and the increasing intensity observed in Fig.\ref{nkallT}c for increasing temperatures is due to the structured ordering of the ice moving from amorphous to a crystalline structure. During heating, the strength of the H-bond network increases through molecular alignment in the ice, and the O$-$O separation between two water molecules decreases \citep{Hagen1981}. This feature also changes the $n$ profile shown in Fig.\ref{nkallT}d. The band at 6~$\mu$m (Fig.\ref{nkallT}e) also provides information about the structure of the H$_2$O ice, namely, the spectral profile is flatter at higher temperatures ($>$~75~K) and narrower at lower temperatures \citep{Hagen1982}. Nevertheless, this band shape variation is not strong in the real refractive indexes (Fig.\ref{nkallT}f). The H$_2$O ice libration band shape and intensity are also sensitive to the temperature variation, which affects the $k$ and $n$ values shown in Fig.\ref{nkallT}g and Fig.\ref{nkallT}h, respectively.

\section{Methodology}

\subsection{Absorption and scattering opacities of H$_2$O ice-coated grains}
\label{opac_sec}
The new H$_2$O refractive index at 30~K presented here differs from previous values in the literature that assumed high $n$ values from experiments performed at high temperatures. Consequently, it is worth quantifying its impact on the derivation of ice-coated grain opacities. To perform this quantification, we construct dust models assuming different geometries, i.e., perfect spheres \citep[Mie theory;][]{Mie1908}, hollow spheres \citep[distribution of hollow spheres - DHS;][]{Min2005} and ellipsoidal grains \citep[continuous distribution of ellipsoids - CDE;][]{Bohren_Huffman}. While the Mie and DHS approaches cover a large grain size distribution, the CDE equations offer an approximation in the Rayleigh limit, i.e., when the grain size is much smaller than the wavelength. Also, we assess the differences between those three geometries and for compact and porous ice.

These calculations are performed using \texttt{optool}, an open-source code dedicated to producing complex dust particle opacities \citep{Dominik2021}. In this study, the grain core composition is the same in all cases, which is amorphous pyroxene \citep[Mg$_{0.7}$Fe$_{0.3}$SiO$_3$;][]{Dorschner1995} and amorphous carbon \citep[][]{Preibisch1993}. The ice mantle is composed of pure water ice and we adopt a wide grain size distribution, ranging from small to large grains, which is a fair approximation to study inter/circumstellar ice analogues. The parameters used in \texttt{optool} to derive the opacities are shown in Table~\ref{opac_par}. The absorption and scattering opacities are given by: 
\begin{subequations}
\begin{eqnarray}
 \kappa_{\rm{abs}} \equiv \frac{C_{\rm{abs}}}{V\rho} 
 \label{eff1}\\
 \kappa_{\rm{sca}} \equiv \frac{C_{\rm{sca}}}{V\rho}
 \label{eff2}
\end{eqnarray}
\end{subequations}
where $C_{\rm{abs}}$ and $C_{\rm{sca}}$ are the absorption and scattering cross sections, $V$ is the volume, and $\rho$ is the density of the icy grain. The cross-section is model-dependent, which comes with different values for different grain geometries, sizes ($s$) and mass fraction (mfrac) of silicate, carbon and ice.

For small grains compared to the wavelength, also called the Rayleigh regime, the cross sections are proportional to the polarizability $\alpha_j$ of spheres and ellipsoids, i.e., $C_{\rm{abs}} \propto \rm{Im}(\alpha_j)$ and $C_{\rm{sca}} \propto |\alpha_j|^2$. The index $j$ represents one of the major axes of the spheroid. Here, the polarizability is given by:
\begin{equation}
    \alpha_j = \frac{V}{4\pi} \frac{\tilde{m_{\rm{av}}}^2-1}{L_j(\tilde{m_{\rm{av}}}^2-1)+1},
\label{plb}
\end{equation}
where $L_j$ is called the depolarization factor and represents the dependence on particle shape. For a sphere, $L_x = L_y = L_z = 1/3$. Similarly, in other geometries such as oblate and prolate spheres, disks and rods, these numbers change and must satisfy the condition $\Sigma_j L_j = 1$. The term $\tilde{m_{\rm{av}}}$ in Equation~\ref{plb} is the combined CRI value of ice and dust, and it shows the dependence of the icy grain cross section with the CRI values. \texttt{Optool} adds the ice mantle to the core by using the Maxwell-Garnett effective medium theory \citet{Garnett1904, Garnett1906}, and $\tilde{m_{\rm{av}}}$ is calculated by: 
\begin{equation}
    \tilde{m}_{\rm{av}}^2 = \tilde{m}_{\rm{core}}^2 \left[1 + \frac{3f(\tilde{m}_{\rm{mantle}}^2 - \tilde{m}_{\rm{core}}^2)/(\tilde{m}_{\rm{mantle}}^2 + \tilde{m}_{\rm{core}}^2)}{1 - f(\tilde{m}_{\rm{mantle}}^2 - \tilde{m}_{\rm{core}}^2)/(\tilde{m}_{\rm{mantle}}^2 + \tilde{m}_{\rm{core}}^2)} \right]
    \label{mav}
\end{equation}
where $\tilde{m}_{\rm{core}}$ and $\tilde{m}_{\rm{mantle}}$ are the CRI values for the dust grain and the ice mantle, respectively; $f$ is the volume fraction of the ice inclusions. 
The effect of the ice porosity is calculated by the following equation obtained by \citep{Hage1990}:
\begin{equation}
    \tilde{m}^2_{\rm{eff}} = 1 + \frac{3(1-P)(\tilde{m}_{\rm{inc}}^2 - 1)/(\tilde{m}_{\rm{inc}}^2 + 2)}{1 - (1 - P)(\tilde{m}_{\rm{inc}}^2 - 1)/(\tilde{m}_{\rm{inc}}^2 + 2)}
\end{equation}
where $P$ is the porosity factor ranging from 0 (i.e., compact icy grain) to 1 - $\delta$ (i.e., porous icy grain), and $\delta \ll 1$. When the ice porosity is considered in the models, $\tilde{m}^2_{\rm{eff}}$ replaces $\tilde{m}^2_{\rm{mantle}}$ in Equation~\ref{mav}.

\begin{table}
\caption{Input parameters used to calculate the absorption and scattering opacities.}
\label{opac_par}      
\centering
\setlength{\tabcolsep}{3pt} 
\renewcommand{\arraystretch}{1.5} 
\begin{tabular}{l c c c}        
\hline\hline                 
\multicolumn{4}{c}{Grain size}\\
\hline
Parameters & Values & ... & ...\\
\hline
$s_{\rm{min}}^{\rm{Mie, DHS}}$ ($\mu$m) & 0.050 & ... & ...\\
$s_{\rm{max}}^{\rm{Mie, DHS}}$ ($\mu$m) & 3000 & ... & ...\\
$s_{\rm{min}}^{\rm{CDE}}$ ($\mu$m) & 0.010 & ... & ...\\
$s_{\rm{max}}^{\rm{CDE}}$ ($\mu$m) & 0.100 & ... & ...\\
porosity & 0\% (compact) & ... & ...\\
porosity & 25\% (porous) & ... & ...\\
\hline
\multicolumn{4}{c}{Grain composition}\\
\hline
Where & mfrac (\%) & $\rho$ (g cm$^{-3}$) & material\\
\hline
Core (silicate) & 0.725\% & 3.01 & pyr-mg70$^a$\\
Core (carbon) & 0.108 & 1.80 & a-C$^b$\\
Mantle (ice) & 0.167 & 0.94 & H$_2$O ice$^c$ (30~K)\\
\hline
\end{tabular}
\tablefoot{$^a$\citet{Dorschner1995}, $^b$\citet{Preibisch1993}, $^c$ This paper.}
\end{table}

For particles of the same size as the wavelength, the cross-sections become a function of Legendre polynomials and Bessel functions \citep{Mie1908, Bohren_Huffman}. The physical meaning is that when a plane electromagnetic wave interacts with a homogeneous sphere in a vacuum, it causes the sphere to produce a non-isotropic outgoing wave. This wave can be expanded using spherical harmonics. For completeness, we quote the cross-sections from \citet{Bohren_Huffman}:
\begin{subequations}
\begin{eqnarray}
 C_{\rm{sca}} &=& \frac{2\pi}{\nu^2} \sum_{j=1}^{\infty} (2j + 1) \left(|a_j|^2 + |b_j|^2 \right)\\
 C_{\rm{ext}} &=& \frac{2\pi}{\nu^2} \sum_{j=1}^{\infty} (2j + 1) \mathbb{R}\left(a_j + b_j \right)\\
 C_{\rm{abs}} &=& C_{\rm{ext}} - C_{\rm{sca}}
 \label{mie_eq}
\end{eqnarray}
\end{subequations}
where $\nu = 2\pi/\lambda$ is the wavenumber of the incoming wave, $j$ is the expansion index, and $a_j$ and $b_j$ are called scattering coefficients, which are a function of $\tilde{m}$. Evidently, when particles are smaller than the involved wavelength, the Mie theory produces cross-sections as the Rayleigh approach. In the case of particles that are large compared to the wavelength $-$ the geometric-optics limit $-$ the same formalism is used, and the calculations become significantly more expensive.

\subsection{Radiative transfer calculations}
\label{rt_sec}
To fully assess the impact of the icy grain opacities resulting from the new CRI values, we run three-dimensional radiative transfer calculations to calculate the spectral energy distribution (SED) of a star surrounded by icy grains. As a proof of concept, we adopt the water ice CRI values at 30~K. The goal of this approach is to assess the differences in the SED by using different ice-dust opacities introduced in \S \ref{opac_sec}. The simulations are carried out by using the radiative transfer code \texttt{RADMC-3D} \citep{Dullemond2012}. We adopt an axisymmetric density structure and perform a full three-dimensional simulation where we keep the source inclinations at 80$^{\circ}$.

Our simplified model assumes that a forming star is surrounded by a spherical envelope with no disk inside. The envelope density profile at different radii ($r$) is the same as used in previous works \citep[e.g.,][]{Pontoppidan2005, Rocha2015} and is given by:
\begin{equation}
    \rho_{\rm{env}}(r) = \left( \frac{R_{\rm{out}}}{r} \right)^{1.5}
\end{equation}
where $R_{\rm{out}}$ is the outer envelope radius. Table~\ref{env_par} summarizes the parameters used in this model.

\begin{table}
\caption{Input parameters used to calculate the mid-IR water ice refractive index.}
\label{env_par}      
\centering
\setlength{\tabcolsep}{3pt} 
\renewcommand{\arraystretch}{1.5} 
\begin{tabular}{l c c}        
\hline\hline                 
\multicolumn{3}{c}{{\bf RADMC-3D parameters}}\\
\hline
Parameters & Description & Value$^a$\\
\hline
R (R$_{\sun}$) & Stellar radius & 1\\
T (K) & Black-body temperature & 800~K\\
M$_{\rm{env}}^{d}$ (M$_{\sun}$) & Envelope dust mass & $10^{-2}$\\
R$_{\rm{env,in}}$ (AU) & Inner envelope radius & 5\\
R$_{\rm{env,out}}$ (AU) & Outer envelope radius & 1000\\
\hline
\multicolumn{3}{c}{{\bf List of simulations}}\\
\hline
Models & Description & Label\\
\hline
Model 1c & Adopts NK1 values and compact ice & M1c\\
Model 1p & Adopts NK1 values and porous ice & M1p\\
Model 2c & Adopts NK2 values and compact ice & M2c\\
Model 2p & Adopts NK2 values and porous ice & M2p\\
\hline
\end{tabular}
\tablefoot{$^a$Values for a generic spherical envelope with an extincted black-body temperature.}
\end{table}

\section{Results}
\label{Results}
\subsection{Ice-grain opacities}
In this section, we compare the opacity values derived from the NK1 and NK2 refractive indexes, which are those using $n_{700nm} = 1.16$ and $n_{700nm} = 1.32$, respectively. No comparisons are made with \citet{Hudgins1993} and \citet{Mastrapa2009}, because of different experimental settings, such as ice growing rate, and chamber pressure which relate to the presence of background gas contaminants, which may induce systematic effects. Additionally, NK1 and NK2 values cover a broader spectral range compared to the literature. In this sense, the comparison of values derived from the same experimental data offers a more consistent approach.

\begin{figure}
   \centering
   \includegraphics[width=\hsize]{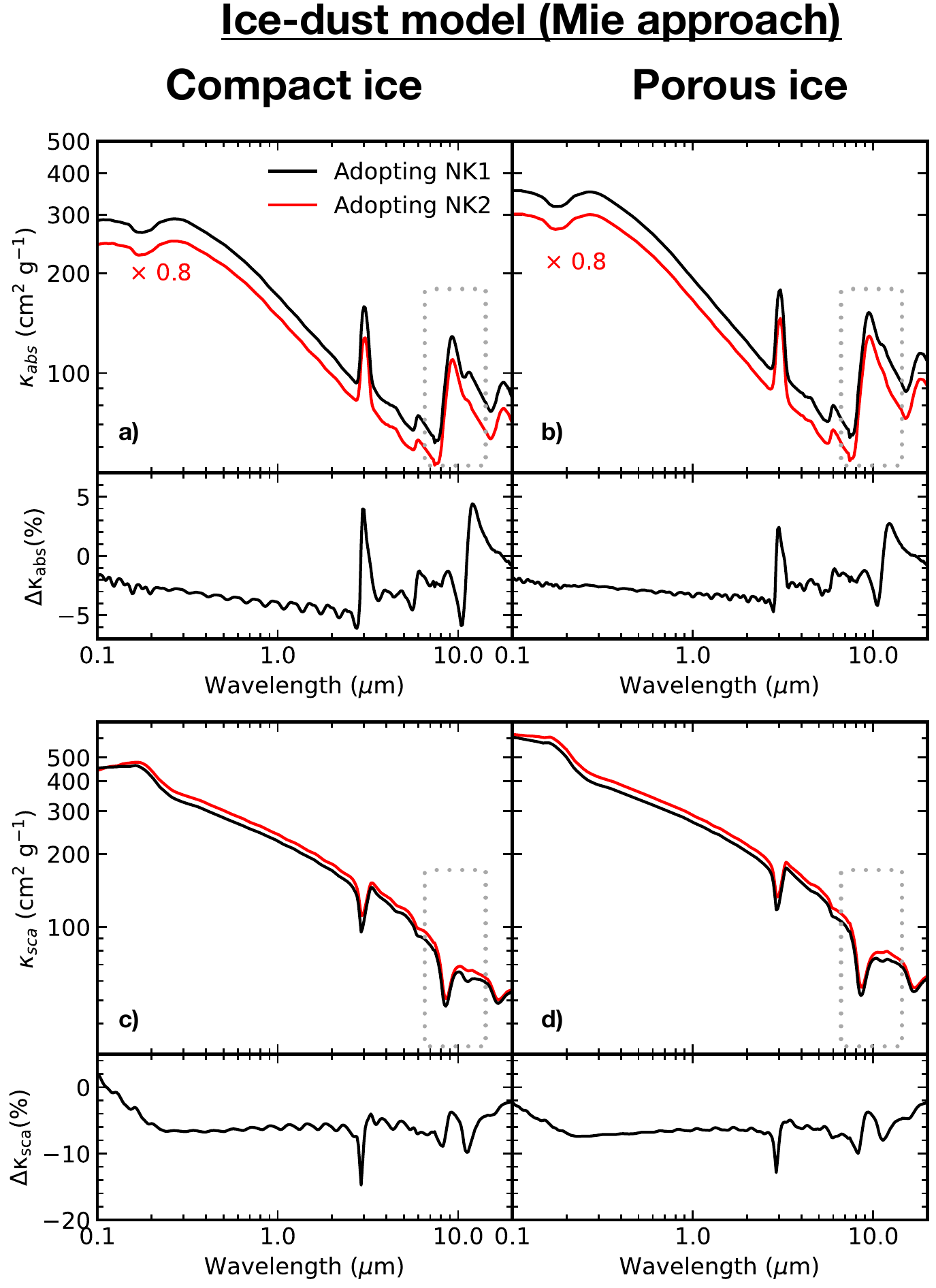}
      \caption{Absorption (upper panels) and scattering (lower panels)  opacities of ice-dust grains assuming compact (left) and porous ices (right). Lines in black are the opacities assuming the water ice refractive index derived in this paper, i.e., assuming $n_{700nm} = 1.16$ (NK1), whereas curves in red show the opacities calculated using $n_{700nm} = 1.32$ (NK2). The small panels below the large ones show the variation between the two opacities in percentage values. A small offset is performed on the absorption opacities for better readability. No offset is applied to the scattering opacity. The rectangular regions indicated by the grey dotted boxes are zoomed-in in Figure~\ref{opac_mie_zoom}.}
         \label{opac_mie}
   \end{figure}

Figure~\ref{opac_mie} shows the absorption and scattering opacities of spherical grains in the framework of the Mie theory, and when compact and porous ices are adopted. The differences between the opacity values calculated from NK1 and NK2 are also shown at the bottom of each panel and are calculated by:
\begin{equation}
    \Delta \kappa_{\rm{abs,sca}} (\%) = \left( \frac{\kappa_{NK1}}{\kappa_{NK2}} - 1 \right) \times 100
    \label{diff_opac}
\end{equation}
where $\kappa_{NK1}$ and $\kappa_{NK2}$ are the opacities derived by assuming $n_{700nm} = 1.16$ and $n_{700nm} = 1.32$, respectively. Quantitatively, the $\Delta \kappa_{\rm{abs,sca}}$ values in panels \ref{opac_mie}{\it a} and \ref{opac_mie}{\it b} show that the absorption opacity calculated from NK1 values reduces the opacity baseline by ~3$-$5\% when compared to the opacities derived from NK2 refractive indexes. Furthermore, it leads to slightly more absorption (4$-$5\%) at wavelengths around $\sim$3~$\mu$m and $\sim$12~$\mu$m, where the O$-$H stretching and the libration bands are located, respectively. Another interesting aspect of using the NK1 values is the increase of absorption opacity around 11.6~$\mu$m, where the H$_2$O libration band is located. Compact ice enhances this feature by 5\% compared to 2.8\% in porous ice. The new refractive indexes derived for NK1 also affect the scattering opacities. Panels \ref{opac_mie}{\it c} and \ref{opac_mie}{\it d} show a reduction in the scattering of around 8\% in most of the wavelengths, and around 13\% at 3~$\mu$m independently whether the ice is compact or porous.

\begin{figure}
   \centering
   \includegraphics[width=\hsize]{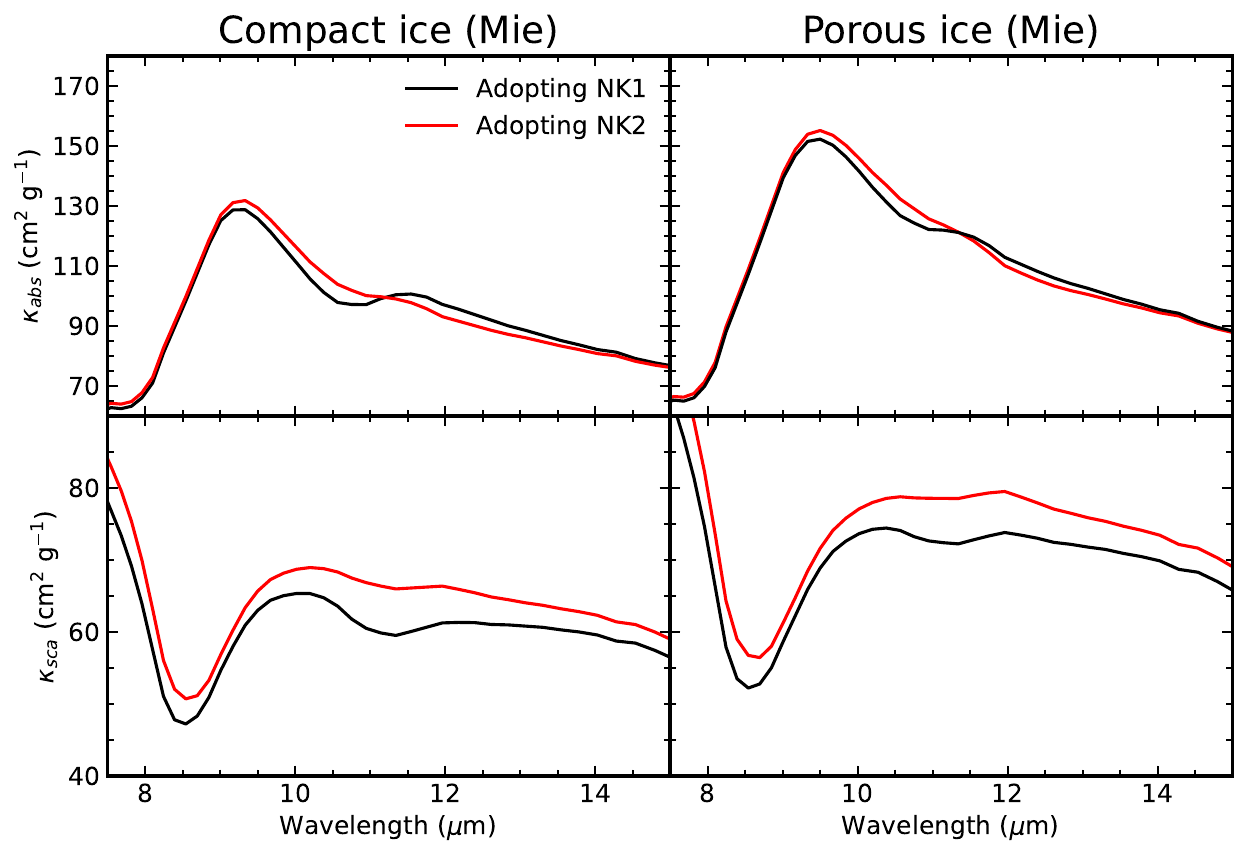}
      \caption{Zoom-in of the absorption and scattering opacities of ice-dust grains shown in Figure~\ref{opac_mie}. The left and right panels show the opacities assuming compact and porous ice grains. No offset is applied in this absorption opacity.}
         \label{opac_mie_zoom}
   \end{figure}

Figure~\ref{opac_mie_zoom} shows zoom-ins between 7.5 and 15~$\mu$m of Figure~\ref{opac_mie}. One can see at 11.6~$\mu$m the presence of the H$_2$O libration band when the NK1 values are used to calculate absorption and scattering opacities. This feature is more prominent when the ice is compact and weaker in porous ice. If NK2 refractive indexes are used, this feature becomes weaker or is no longer visible.  

\begin{figure}
   \centering
   \includegraphics[width=\hsize]{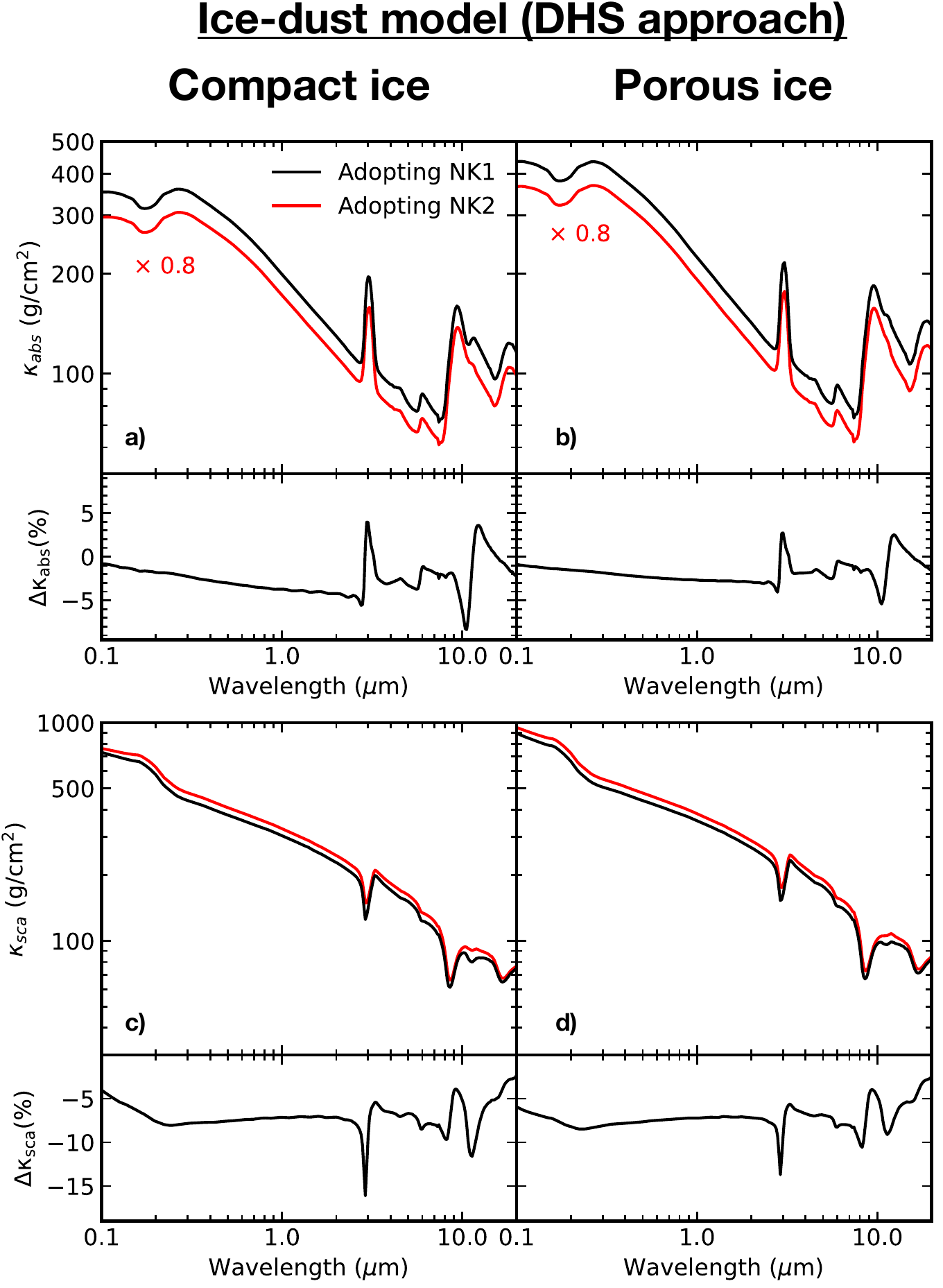}
      \caption{Same as Figure~\ref{opac_mie}, but assuming the DHS approach. An offset on the absorption opacity is used for better readability. No offset is applied on the scattering opacity.}
         \label{opac_dhs}
   \end{figure}

In Figure~\ref{opac_dhs}, we investigate the differences in the opacities for both compact and porous grains when the DHS model is considered. The absorption and scattering opacities shown in panels a$-$d have a similar trend compared to the Mie ice-dust models (Fig. \ref{opac_mie}). There is an increase in the absorption opacity of 3$-$5\% around 3~$\mu$m and 12~$\mu$m. Additionally, the bump around 11.6~$\mu$m associated with the H$_2$O libration band is also present in the model of compact ices. Conversely, this band is hidden when NK2 values are adopted to derive the opacities or when the ice is porous. Again, the most notable effect of using NK1 values is seen in the scattering opacities. Specifically, around 3~$\mu$m and 12~$\mu$m, the scattering is reduced by around 15\% and 10\%, respectively.

\begin{figure}
   \centering
   \includegraphics[width=\hsize]{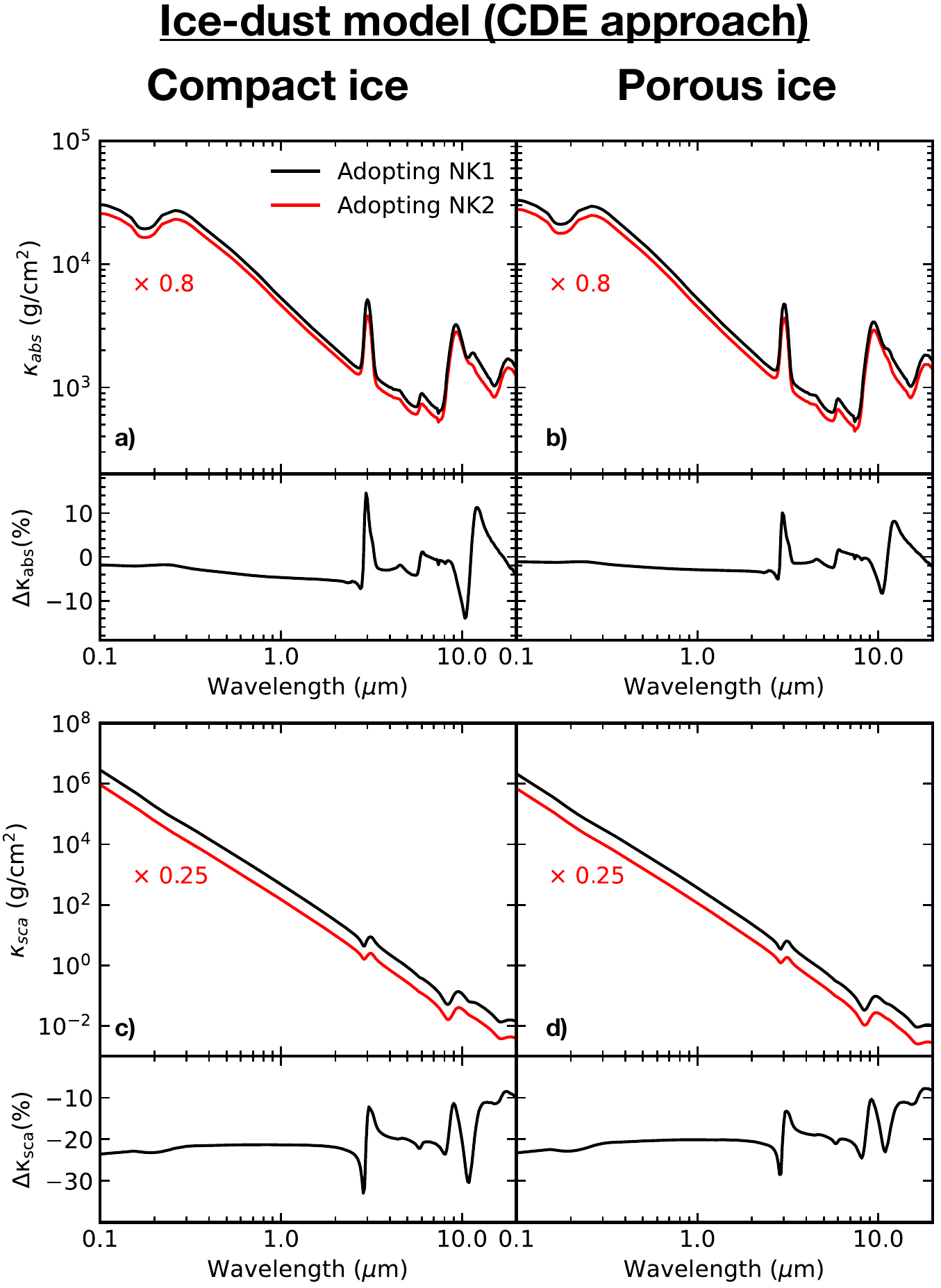}
      \caption{Same as Figure~\ref{opac_mie}, but assuming the CDE approach. An offset on both absorption and scattering opacities is applied for better readability.}
         \label{opac_cde}
   \end{figure}

Finally, in Figure~\ref{opac_cde} we assess the impact of the new H$_2$O ice refractive index on grains modelled using the CDE approach, i.e., assuming grains with ellipsoidal shapes. First, we note the increase in the absorption opacity around 3~$\mu$m and 12~$\mu$m of $\sim$12\%, which is twice higher than in the Mie and DHS models. Second, the H$_2$O libration band is a persistent feature in compact ices when NK1 refractive indexes are used. As in the other cases, it becomes weaker again when the ice is porous or when the NK2 values are considered. Moreover, the NK1 values have a substantial impact on the scattering opacity. Specifically, it is reduced by as much as 28-34\% at 3~$\mu$m.

\begin{figure}
   \centering
   \includegraphics[width=\hsize]{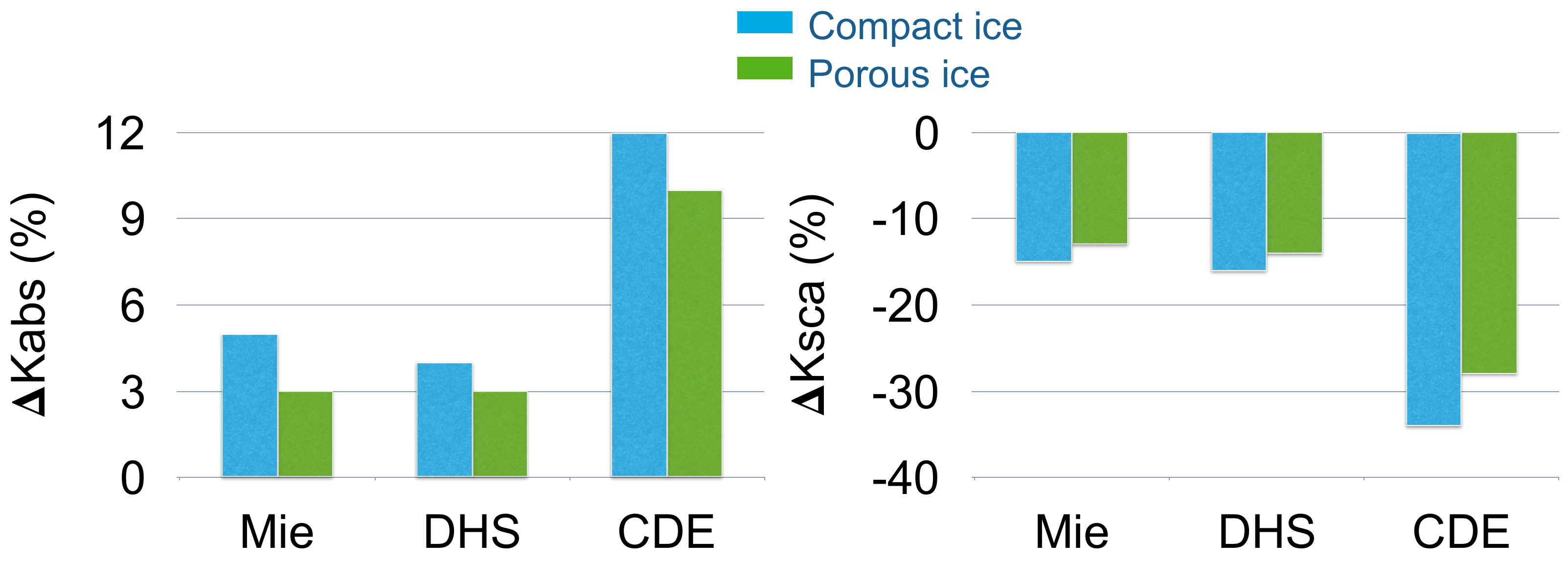}
      \caption{Bar plot showing the percentage difference between absorption (left) and scattering (right) opacities calculated from NK1 and NK2 values for compact (blue) and porous ice (violet). The bars are grouped by the method adopted to calculate opacities, i.e., Mie, DHS and CDE. Positive and negative values indicate an increase and decrease in the opacity value, respectively. }
         \label{gmodels}
   \end{figure}

The differences in the absorption and scattering opacities around 3~$\mu$m shown in Figures~\ref{opac_mie}$-$\ref{opac_cde} are summarized in Figure~\ref{gmodels}. We note that (i) there are more significant differences in the opacities when the ice is compact, (ii) opacity differences in CDE models are more prominent than in Mie and DHS models, (iii) the new H$_2$O ice refractive index reduces by more than 25\% the scattering opacity in CDE models, (iv) the differences between Mie and DHS models are comparable. This stresses the need for accurate refractive indexes of the most abundant molecules in ice when using them to derive opacities for spectral data interpretation.

\subsection{Protostellar spectrum and H$_2$O ice column density}
\begin{figure}
   \centering
   \includegraphics[width=\hsize]{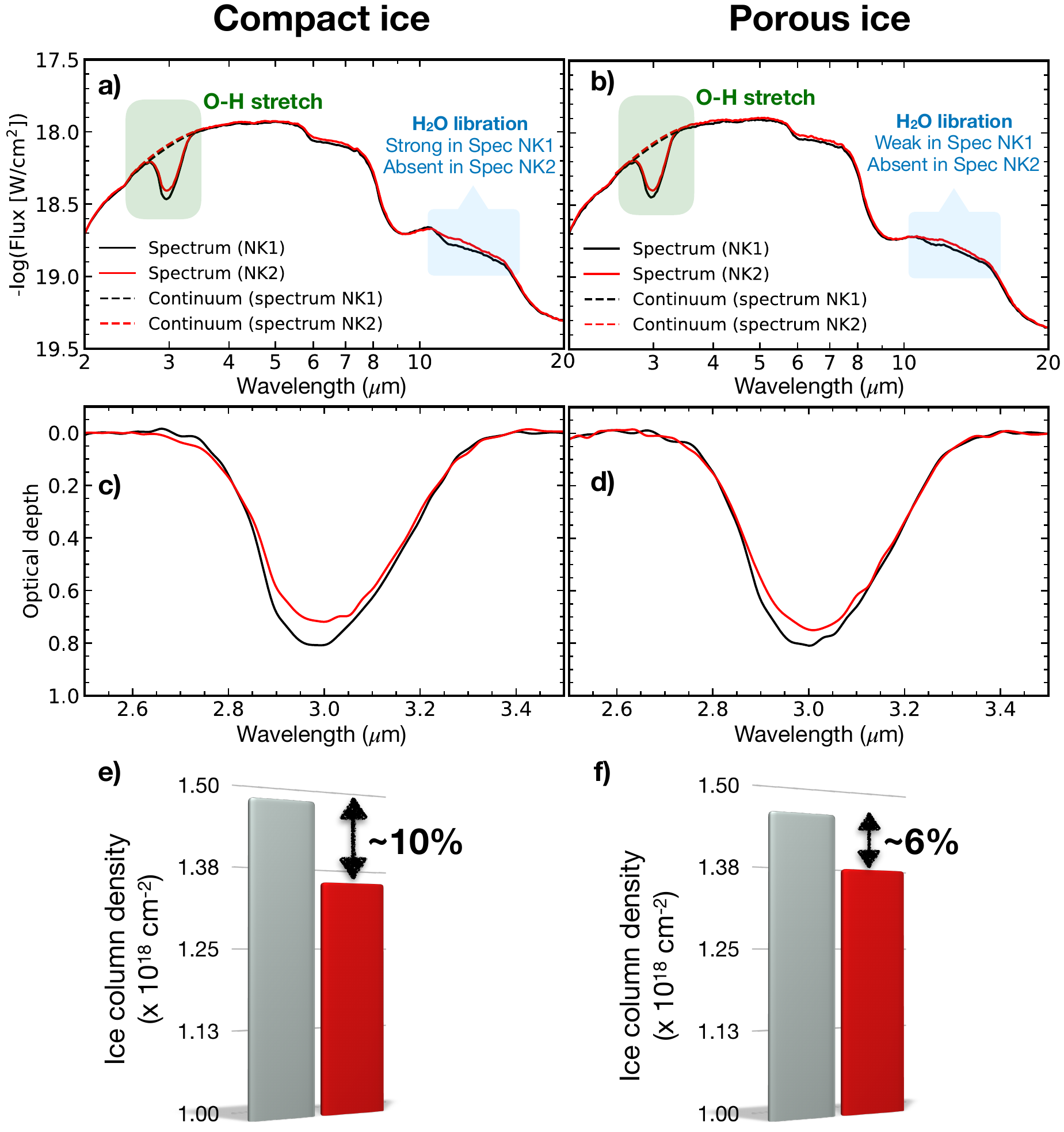}
      \caption{Effects of opacity values derived for grains under the DHS approach by assuming the NK1 and NK2 values. Panel~$a$ shows the synthetic protostellar spectrum with H$_2$O ice and silicate absorption bands calculated from opacity models based on NK1 (black) and NK2 (red) refractive index values. The black and red and dashed lines over the 3~$\mu$m feature are the continuum. The blue box around 13~$\mu$m highlights the absence of the H$_2$O libration band in the spectrum derived from NK2 values. Panel~$b$ displays the same as panel~$a$ but for porous ice. Panels~$c$ and $d$ present the H$_2$O ice column density derived from the optical depth spectrum using compact and porous ices, respectively. Finally, Panels~$e$ and $f$ compare the ice column densities from both cases (grey: NK1; red: NK2).}
         \label{Nice}
   \end{figure}

In order to make the previous findings more concrete, we run radiative transfer models using the \texttt{RADMC-3D} code to compute the effect of the ice opacities in the spectrum of a protostar surrounded by an envelope. Specifically, we use opacities derived from models M1c, M1p, M2c and M2p (see Table~\ref{env_par}) under the DHS and CDE approaches. Figures~\ref{Nice}a and \ref{Nice}b show the mid-IR spectrum of grains under the DHS approach. In models M1c, M1p, silicate and all H$_2$O ice bands are seen, including the bump around the water ice libration mode at 11.6~$\mu$m. In models M2c and M2p, the absorption excess at 11.6~$\mu$m is not visible in the synthetic spectra as indicated by the opacity values.

The water ice bands have different absorption intensities in models M1c and M1p. We use these synthetic protostellar spectra to assess the impact of the NK1 and NK2 values on the determination of the H$_2$O ice column density. Figures~\ref{Nice}c and \ref{Nice}d compare the optical depth spectrum of the H$_2$O ice band at 3~$\mu$m in both cases. To convert the spectrum from flux to optical depth scale we use the equation $\tau_{\lambda} = -\rm{ln}(F_{\lambda}^{obs}/F_{\lambda}^{cont})$, where $F_{\lambda}^{obs}$ is the protostellar flux and $F_{\lambda}^{cont}$ is the continuum flux. The continuum is determined by a low-order polynomial function. We note that the NK1 refractive index values lead to more light extinction at 3~$\mu$m both when the ice is porous and compact (Models M1p and M1c), compared to the models considering the NK2 values (M2p and M2c). Moreover, they result in slightly more absorption if the ice is compact instead of porous.   

Finally, we derive the H$_2$O ice column densities from the O$-$H stretching mode at 3~$\mu$m. Figures~\ref{Nice}e and \ref{Nice}f display bar plots to compare the differences between the ice column densities. These are calculated with the following equation, $N_{\rm{ice}} = 1/A \int_{\lambda} \tau_{\lambda} d\lambda$, where $A$ is the band strength of the O$-$H stretching mode, i.e., $2 \times 10^{-16}$ cm/molecules \citep{Gerakines1995}. We can see from panel~\ref{Nice}e that if the ices are compact, the column density is 10\% higher for NK1 values, compared to NK2 values. Panel~\ref{Nice}f shows that if the ices are porous, the difference between column densities is around 6\%. This underlines, once again, that in order to interpret astrophysical observables, such as column densities, precise refractive index values are an absolute prerequisite.

\begin{figure}
   \centering
   \includegraphics[width=9cm]{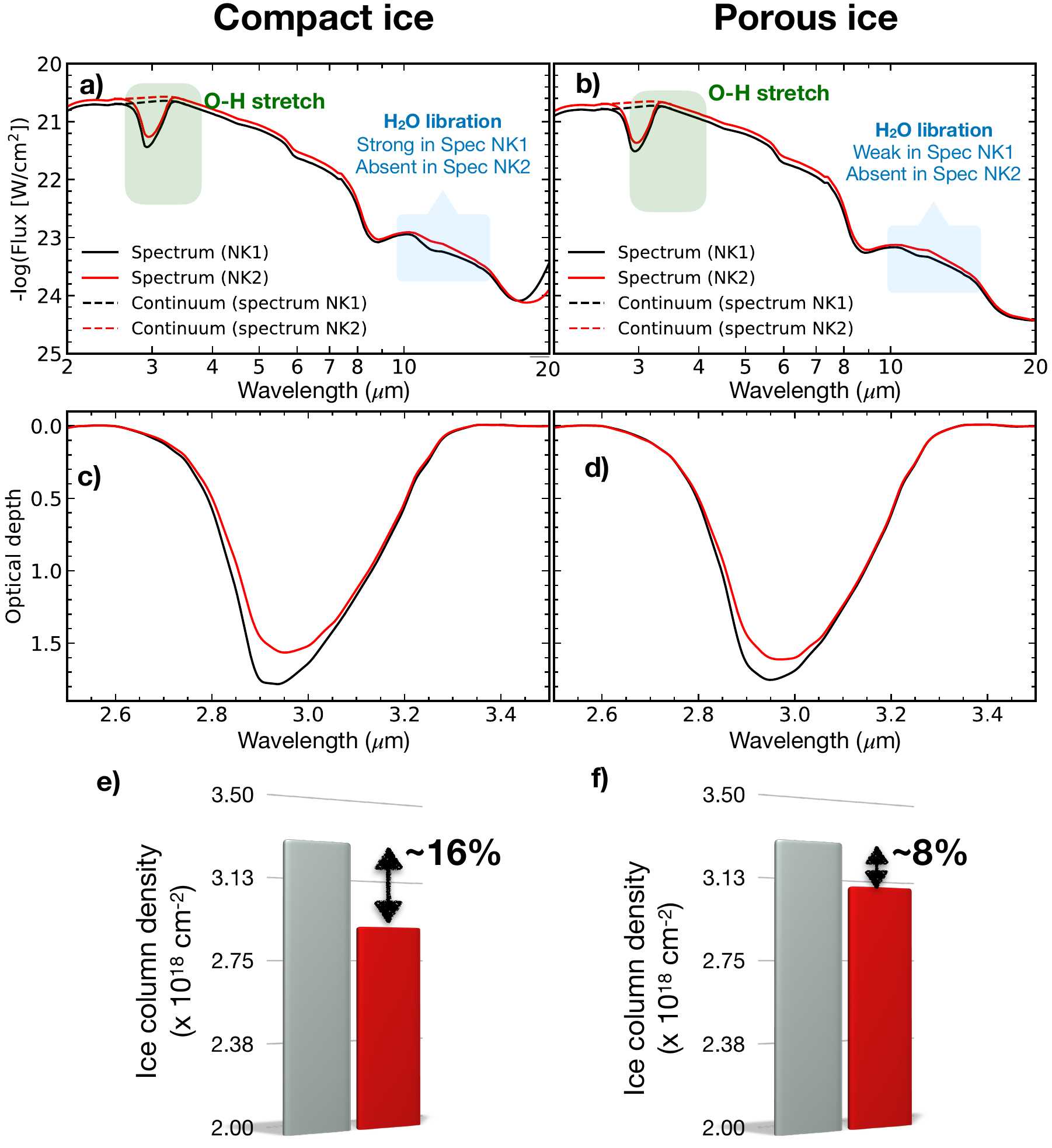}
      \caption{Same as Figure~\ref{Nice}, but assuming icy grains under CDE approach.}
         \label{Nice_cde}
   \end{figure}

We repeat this analysis on the protostar synthetic spectra when the opacity values are derived from the CDE models, and the result is shown in Figure~\ref{Nice_cde}. We note that the differences in the water ice column densities are bigger. In the case of compact grains, the H$_2$O ice column density is 16\% higher than in the cases where NK2 values are used.

\section{Discussion}
We have shown that the NK1 and NK2 refractive index values impact the derivation of physical parameters such as the scattering and absorption opacities, and the ice column density derivation from radiative transfer models. Now, we discuss how these results may be affected by different initial conditions, such as the effect of porosity, ice mass, and grain size on the astrophysical observables.

\subsection{H$_2$O ice libration band and ice mass}
\begin{figure}[h!]
   \centering
   \includegraphics[width=8cm]{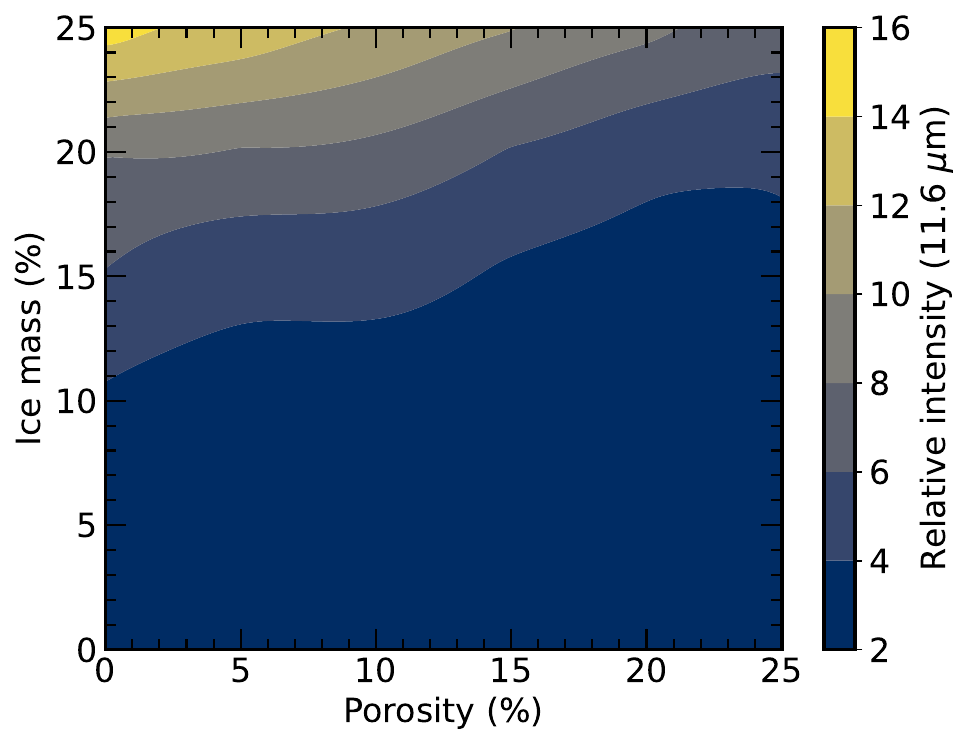}
      \caption{Relative intensity of the 11.6~$\mu$m profile by assuming different ice mass and porosity.}
         \label{degen}
   \end{figure}

The persistent presence of the H$_2$O libration band in compact ice models using Mie, CDE and DHS approaches shows that this feature is independent of the dust geometry assumed to derive the opacities. It is rather associated with the amount of ice (ice mass) adopted in the models, which is given in percentage relative to the total dust core mass. In Figure~\ref{degen}, we explore how the relative intensity of the band at 11.6~$\mu$m changes as a function of the ice porosity and mass. The relative intensity refers to the opacity value after local baseline subtraction around 11.6~$\mu$m. The porosity and ice mass vary from 0 to 25\%. From this figure, we find that the highest intensity (light yellow colour) is when the ice is compact (no porosity) and the ice mass is 25\% of the total dust mass. The dark blue colour indicates the lowest relative intensity. It should be noted that this figure does not constrain by itself the detection limit to observe the 11.6~$\mu$m since this depends on the protostar and on the sensitivity of the instruments used in the observations.


\begin{figure}[h!]
   \centering
   \includegraphics[width=8cm]{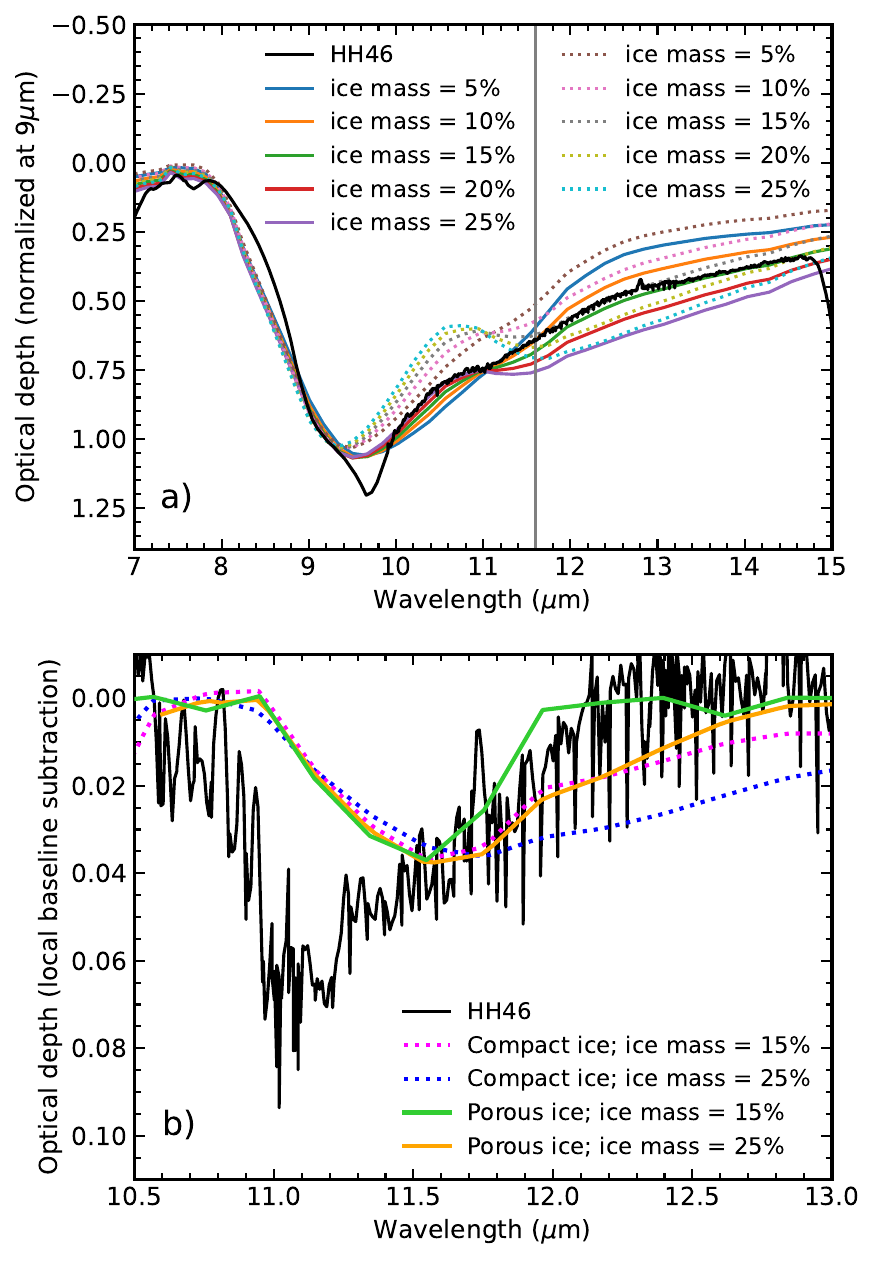}
      \caption{a) Normalized synthetic spectra in optical depth scale of models assuming compact (coloured dotted lines) and porous (coloured solid lines) ice and different ice mass. The spectrum of HH~46 is shown in black for comparison. The vertical grey line indicates the profile at 11.6~$\mu$m. b) Comparison between the observed profile around 11.6~$\mu$m in HH~46 with four synthetic spectra after local baseline subtraction. The synthetic spectra are derived by using two ice masses (15\% and 25\% relative to dust mass) for compact and porous ice.}
         \label{lib}
   \end{figure}

\begin{figure}[h!]
   \centering
   \includegraphics[width=8cm]{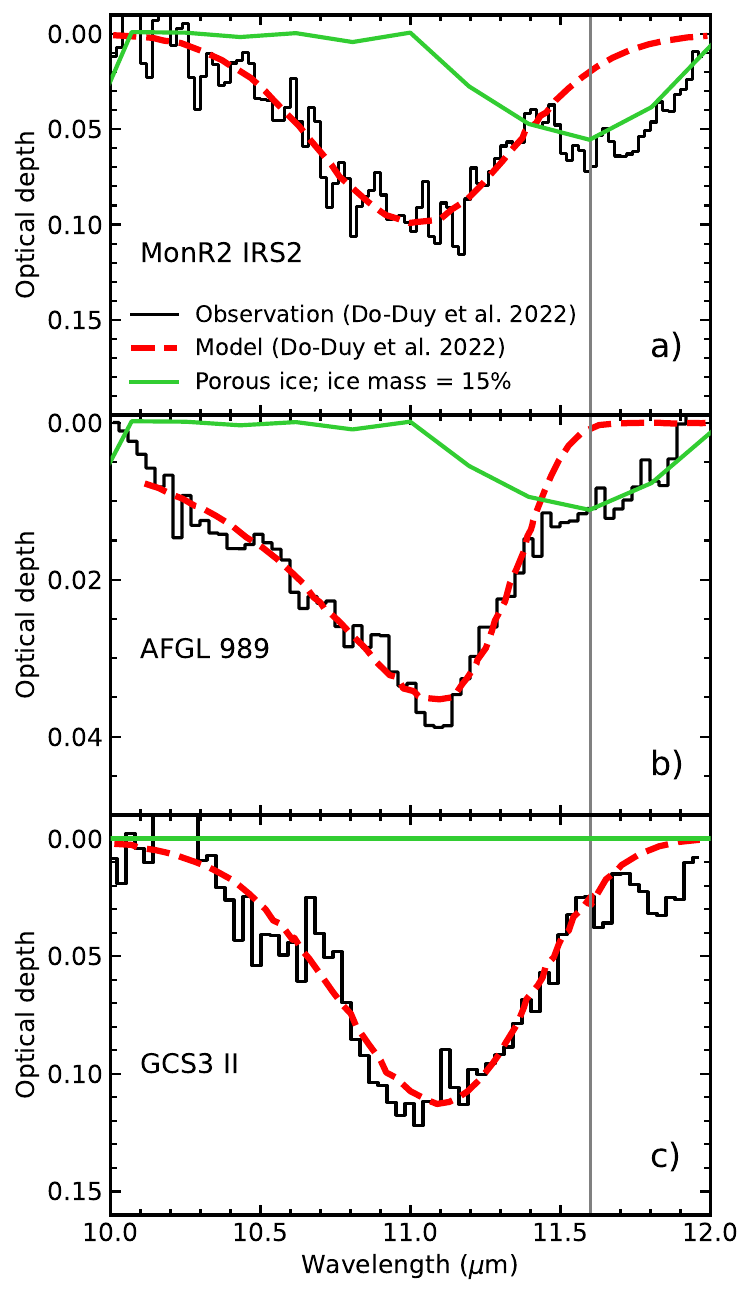}
      \caption{Comparison between the absorption feature around 11~$\mu$m observed towards different sources with the H$_2$O ice libration band from synthetic spectra. Panels a, b and c  show in black the profiles for the high-mass protostars Monoceros R2-IRS, AFGL~989 and the galactic centre source GCS3~II, respectively. The dashed red line is an asymmetric Gaussian fit of the 11.2~$\mu$m performed by \citet{DoDuy2020}. The solid green line is the profile of the H$_2$O ice libration band from a model assuming an ice mass of 15\% distributed in porous ice and scaled to the peak optical depth at 11.6~$\mu$m, which is indicated by the vertical grey line.}
         \label{lib_silic}
   \end{figure}

Figure~\ref{lib_silic} shows additional data that suggests the presence of porous water ice toward high-mass protostars. The absorption profile around 11~$\mu$m is taken from \citet{DoDuy2020} which is obtained after local baseline subtraction. An asymmetric Gaussian profile is used to fit the crystalline silicate feature. Towards two high-mass sources, MonR2~IRS2 and AFGL~989, there is an absorption excess at 11.6~$\mu$m that is not fitted with the Gaussian profile. However, this band shows a good match with the scaled H$_2$O ice libration band obtained from the simulated protostar spectrum considering an ice mass of 15\% and porous structure, which is similar to HH46 (see Figure~\ref{lib}). The two protostars are known to have H$_2$O ice detected at 3 and 6~$\mu$m \citep{Gibb2004}, which further supports the presence of water ice at 11.6~$\mu$m. Conversely, GCS3 does not have H$_2$O ice detection associated with it, and its silicate profile is often used as a template to remove the silicate features towards other protostars \citep[e.g.,][]{Boogert2008, Bottinelli2010, Rocha2021}. In fact, no water ice profile is suggested from this comparison.

Typically, laboratory studies consider the OH dangling bond of H$_2$O ice at 2.7~$\mu$m as a tracer of the ice porosity \citep[e.g.,][]{McCoustra1996, Palumbo2005, Nagasawa2021}, although the absence of this band is not proof the ice is fully compact \citep{Isokoski2014, Bossa2015}. Until now, the OH dangling bond was tentatively observed only in the background star NIR38 with JWST under the Ice Age ERS program \citep{McClure2023}. Earlier astronomical observations were unable to detect this band, likely because of limitations in instrumental sensitivity. Moreover, the intensity of the OH dangling mode in the near-infrared range may be severely reduced due to the circumstellar dust extinction.

In addition to the OH dangling bond, the comparisons shown in Figures~\ref{lib} and \ref{lib_silic}, indicate that the H$_2$O libration band profile can be used as a diagnostic of the ice porosity in protostellar environments, as well. In this regard, we encourage further work around the silicate feature at 9.8~$\mu$m to constraints on the ice structure. Specifically, low-mass protostars are ideal targets to be observed with JWST, and the sensitivity of the Mid-Infrared Instrument (MIRI) allows for achieving this goal.

\subsection{How big are the grains to hide the H$_2$O ice feature in the mid-IR?}
The improved H$_2$O ice CRI values derived from this work also extend the discussion on how the size of the grains affects the strength of the water ice absorptions at 3 and 6~$\mu$m. Based on the study of water ice in the coma of comet Hale-Bopp, \citet{Lellouch1998} estimated a mean icy grain size of 15~$\mu$m to hide the IR water ice feature at 3~$\mu$m. However, the authors indicated that their assumption of a single-grain size is not realistic for sublimating icy grains. Additional work by \citet{van_Dishoeck2021} calculated a grain size threshold of 10~$\mu$m for the detection of the 3~$\mu$m band.

\begin{figure}[h!]
   \centering
   \includegraphics[width=\hsize]{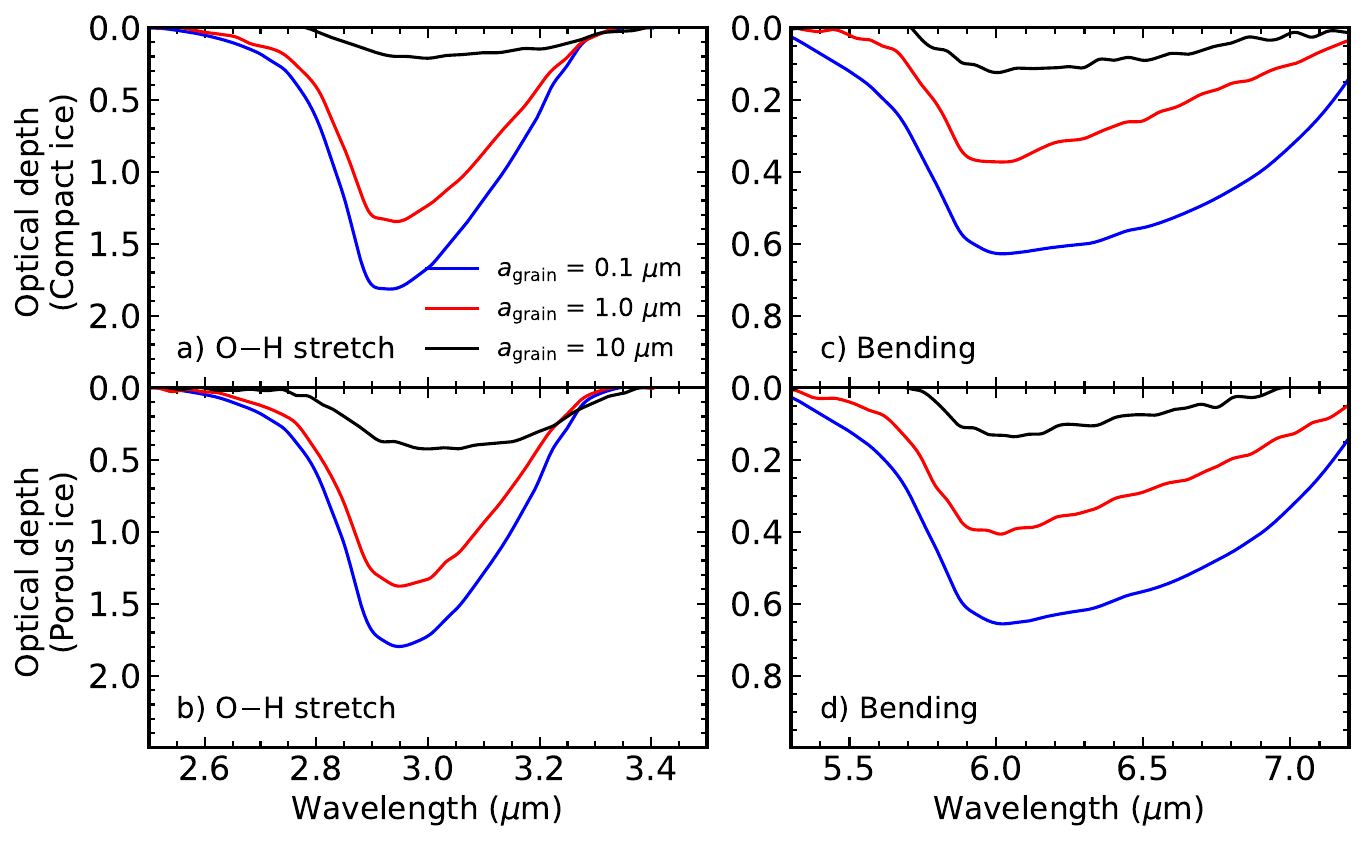}
      \caption{Optical depth variation with grain growth and ice porosity. Panels a and b show the changes for the O$-$H stretching mode, and panels c and d show the H$_2$O ice bending mode. Note the different vertical scales for the two wavelength regimes.}
         \label{od_gsize}
   \end{figure}

In Figure~\ref{od_gsize}, we show the changes in the optical depth at 3 and 6~$\mu$m when the grain size increases and different porosity levels are taken into account. For these calculations, we chose the DHS approach that simulates irregular grain shapes and we keep an ice mass of 25\% constant. For compact ices shown in panels {\it a} and {\it b}, increasing the grain size from $a_{\rm{grain}} = 0.1~\mu$m to $a_{\rm{grain}} = 1~\mu$m, reduces the ice column density by $\sim$33\%. This reduction is more significant, i.e., a factor of 8 when the grain grows to $a_{\rm{grain}} = 10~\mu$m. For porous ices, the reduction in the column density is 30\% and a factor of 4 if the grain grows from 0.1~$\mu$m to 1~$\mu$m and 10~$\mu$m, respectively. Notably, the difference is lower when the ice is porous. We perform the same analysis on the 6~$\mu$m band, shown in panels {\it c} and {\it d}. The ice column density decreases by 40\% and 92\% when the icy grain grows from 0.1~$\mu$m to 1~$\mu$m, and from 0.1~$\mu$m to 10~$\mu$m, respectively. When the ice is porous, the reduction is 35\% and 80\%, respectively.


\begin{figure}
   \centering
   \includegraphics[width=\hsize]{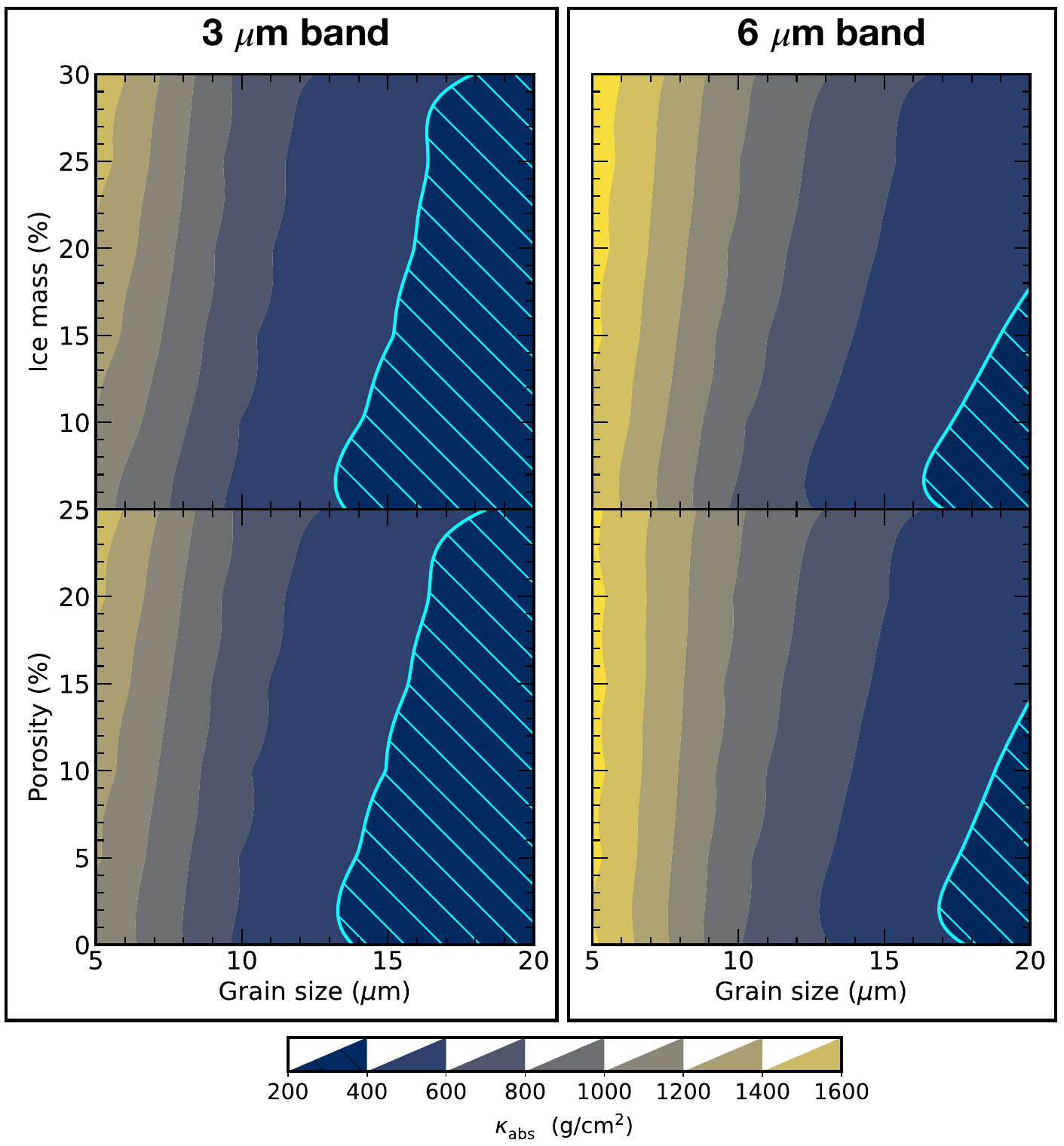}
      \caption{Absorption opacity map of the 3~$\mu$m (left) and 6~$\mu$m (left) water ice bands for different grain sizes, ice porosity and ice mass. The blue hatched area indicates the region where the H$_2$O ice modes are no longer detectable.}
         \label{degen3mic}
   \end{figure}

Based on the variations in the intensities of the H$_2$O ice bands with grain size, we explore the combined effect of grain growth and ice mass and porosity variation. In Figure~\ref{degen3mic}, we show a contour map of absorption opacities at 3~$\mu$m (left) and 6~$\mu$m (right) by changing porosity and ice mass. The blue hatched area indicates an optical depth below 0.01, i.e., the ice band is no longer observable. At 3~$\mu$m, the O$-$H stretching mode can be detected for grains with sizes from 13~$\mu$m to 18~$\mu$m, depending on the ice mass and porosity. This occurs because the ice mass is kept constant when the size of the grain increases. Therefore the amount of ice distributed on the grain surface decreases (see Equations~\ref{eff1} and \ref{eff2}). We perform the same analysis of the H$_2$O bending mode at 6~$\mu$m. The minimum grain size threshold is around 16$-$17~$\mu$m, but this water feature in grains larger than 20~$\mu$m remains observable because the opacity is lower at this wavelength. 

An interesting side effect of the work discussed here is that this analysis may also help to further constrain the problem of the missing oxygen, i.e., the lower oxygen budget in star-forming regions compared with diffuse clouds. An elegant solution proposed for cold protostar envelopes is that large grains suppress H$_2$O ice features, which are not observable. In our analysis, grains need to grow larger than 10~$\mu$m and 15~$\mu$m to suppress the 3~$\mu$m and 6~$\mu$m water bands, respectively. Although it remains a challenge how to explain grain growth at the level of mm-size via coagulation process \citep[e.g.,][]{Silsbee2022}, there are sufficient observational studies showing that $\mu$m-size grains are rather plausible in dense clouds \citep[e.g.,][]{Pagani2010, Miotello2014, Dartois2022}. It is important to highlight that there are other oxygen-bearing molecules hosting a large amount of oxygen in the solid phase, and that are not easily visible in the infrared spectrum. For example, molecular oxygen (O$_2$), has not been detected yet in interstellar ices. New experimental works have explored this problem further and provided laboratory data to search for O$_2$ in ices \citep{Muller2022}. 

We also note that pure H$_2$O ice does not exist in space; it is typically mixed with other hydrogenated species \citep[e.g., NH$_3$, CH$_4$;][]{Dartois2001, Oberg2008, Qasim2018}, as well as with CO, CO$_2$, CH$_3$OH \citep{Gerakines1999, Pontoppidan2008, McClure2023} after the CO freeze-out. Nevertheless, \citet{Isokoski2014} shows that the real refractive index at 632.8~nm of the H$_2$O:CO$_2$ ice mixture at 20K is 1.22 for a CO$_2$ ice fraction of 10\% and 1.24 for a fraction of 20\%. These numbers are around 5\% higher than the $n_{700nm}$ used in this paper for the pure H$_2$O ice, but still 10\% lower than the typical value used in the literature (1.32). Despite this caveat, the numbers discussed here definitely will be useful for studying low-temperature environments dominated by water ice. In the nearby future, new work will provide temperature and wavelength-dependent CRI values for ice mixtures, to improve on recording data of astrophysical relevance, as well as to further explore the role of the ice mixtures in the interpretation of ice observations.

\section{Conclusions}
\label{conclusion}
We show how recently derived broad band complex refractive index values, using a new technique, impact astrophysical observables, such as ice column density and grain size. For this, we have studied in detail the UV- to mid-IR complex refractive index of H$_2$O ice for a large range of temperatures 30, 75, 105, 135 and 160~K. Below 160~K, the CRI values are lower than previously estimated in the literature. At 30~K, the real refractive index is around 14\% lower. Below we summarize the main conclusions of this work:

\begin{itemize}

    \item The new refractive index values of H$_2$O ice-covered grains lead to more absorption and less scattering for the 3~$\mu$m band. Quantitatively, this difference is due to the grain model adopted, and whether the ice is porous or compact. Under Mie and DHS approaches, the values are similar and range from 3$-$5\% more absorption and 15\% less scattering. Under the CDE approach, there is around 10$-$12\% more absorption and 28-35\% less scattering.

    \item Ice column densities derived from simulated spectra using the new and previous refractive indexes are different. Under the CDE approach, the H$_2$O ice column density derived from the 3~$\mu$m band is 16\% and 8\% higher if the ice is compact and porous, respectively. Under the DHS approach, the differences are 10\% and 6\%, respectively.

    \item Adopting the new refractive index values to derive grain opacities suggests that the H$_2$O has a prominent feature at 11.6~$\mu$m, which is not visible when previous refractive indexes are used. The shape of this band changes with the ice porosity level. We find that when this band is associated with a porosity level of 15\%, it can explain the absorption excess observed toward the low-mass protostar HH~46 and the high-mass protostars MonR2~IRS2 and AFGL~989. Therefore, in addition to the OH dangling mode at 2.7~$\mu$m, the water libration band offers an additional spectral range to constrain the porosity level of interstellar ices.

    \item The grain size threshold to detect the 3~$\mu$m feature is pushed further in this work to 18~$\mu$m and depends on the ice mass and porosity. Likewise, the 6~$\mu$m band remains detectable even for grains larger than 20~$\mu$m.

\end{itemize}

\begin{acknowledgements}
We thank the referee, James Stubbing, for the constructive comments that improved the quality and clarity of this manuscript. This project has received funding from the European Research Council (ERC) under the European Union’s Horizon 2020 research and innovation programme (grant agreement No. 291141 MOLDISK). This work is supported by the Danish National Research Foundation through the Center of Excellence “InterCat” (Grant agreement no.: DNRF150), and by the Netherlands Research School for Astronomy (NOVA), grant 618.000.001. WRMR also thanks Ardjan Sturm for stimulating discussions, Carsten Domenik for making the \texttt{Optool} code publicly available. This research has made use of NASA's Astrophysics Data System Bibliographic Services (ADS).          
\end{acknowledgements}

\bibliographystyle{aa}
\bibliography{References}


\appendix

\end{document}